\newif\ifmnras
\mnrastrue
\ifmnras
	\documentclass[a4paper,fleqn,usenatbib]{mnras}
\else
	\documentclass[apj,numberedappendix]{emulateapj}
\fi

\usepackage{graphics,epsf}
\usepackage{amsmath}                % American Mathematical Society package
\usepackage{amsfonts}               % American Mathematical Society fonts
\usepackage{amssymb}                % American Mathematical Society symbol
\usepackage{epsfig}                 % EPS figures
\usepackage{graphicx}
\def \msyr{~\rm{M_{\odot}}~\rm{yr^{-1}}}
\def \cm{~\rm{cm}}
\def \s{~\rm{s}}
\def \km{~\rm{km}}

\def \g{~\rm{g}}

\def \AU{~\rm{AU}}

\def \days{~\rm{day}}

\def \rmModot{~\rm{M_\odot}}

\ifmnras
	\def \aap{A\&A}
	
	\def \apj{ApJ}
	\def \apjl{ApJ}
	\def \apjs{ApJS}

	\def \mnras{MNRAS}

\fi

\usepackage{xcolor}
\definecolor{redak}{rgb}{0.9,0.15,0.05}
\ifmnras
	\title[3D simulations of GEE]{Simulating a binary system that experiences the grazing envelope evolution}

	\author[S. Shiber and N. Soker]{
Sagiv Shiber,$^{1}$\thanks{E-mail: \href{shiber@campus.technion.ac.il}{shiber@campus.technion.ac.il}}
Noam Soker$^{1}$\thanks{E-mail: \href{mailto:soker@physics.technion.ac.il}{soker@physics.technion.ac.il}}
\\
$^{1}$Physics Department, Technion -- Israel Institute of Technology, Technion City -- Haifa 3200003, Israel\\
}

	\date{Accepted XXX. Received YYY; in original form ZZZ}

	\pubyear{2017}
\fi

\begin{document}
\label{firstpage}

\ifmnras
	\pagerange{\pageref{firstpage}--\pageref{lastpage}}
	\maketitle
\else
	\title{Simulating a binary system that experiences the  grazing envelope evolution}

	\author{Sagiv Shiber}
	\author{Noam Soker}
	\affil{Department of Physics, Technion -- Israel
	Institute of Technology, Haifa 32000, Israel;
	shiber@campus.technion.ac.il; soker@physics.technion.ac.il}
\fi

\begin{abstract}
We conduct three-dimensional hydrodynamical simulations, and show that when a secondary star launches jets while performing spiral-in motion into the envelope of a giant star, the envelope is inflated, some mass is ejected by the jets, and the common envelope phase is postponed. We simulate this grazing envelope evolution (GEE) under the assumption that the secondary star accretes mass from the envelope of the asymptotic giant branch (AGB) star and launches jets. In these simulations we do not yet include the gravitational energy that is released by the spiraling-in binary system. Neither do we include the spinning of the envelope. Considering these omissions, we conclude that our results support the idea that jets might play a crucial role in the common envelope evolution, or in preventing it.  
\end{abstract}

\begin{keywords}
binaries: close --- stars: AGB and post-AGB --- stars: winds, outflows --- ISM: jets and outflows
\end{keywords}

% ==========================================================
\section{INTRODUCTION}
\label{sec:intro}
% ==========================================================

Jets that are launched by a compact object that accretes mass from an extended ambient gas play significant roles in influencing back the ambient gas in many types of astrophysical objects. In most cases the system operates in a negative
feedback cycle, the jet feedback mechanism (JFM; \citealt{Soker2016Rev}). 
 
One such case might be the common envelope evolution. 
\cite{ArmitageLivio2000} and \cite{Chevalier2012} discuss the ejection of the common envelope by jets that are launched from a neutron star that is spiraling-in inside a giant envelope. They, however, do not consider jets to play a role when the compact companion of the common envelope evolution is not a neutron star. 
We here follow earlier suggestions that jets play a significant role in removing the common envelope when other compact companions are considered as well, in particular main sequence companions (e.g., \citealt{Soker2004, Papishetal2015, Soker2015, Soker2016Rev, MorenoMendezetal2017}). 

We do note that in a recent paper \cite{Murguiaetal2017} argue that disk formation in common envelope evolution is rare, and so is the JFM in common envelope evolution. On the other hand, there are arguments that the accreted gas can launch jets even when the angular momentum is below the Keplerian angular momentum on the surface of the mass-accreting compact companion \citep{Shiberetal2016, SchreierSoker2016}. 
In addition, \cite{BlackmanLucchini2014} study the momenta of the outflow of bipolar pre-planetary nebulae, and conclude that strongly interacting binary systems launch energetic jets, probably through a common envelope evolution. 
 In any case, \cite{Murguiaetal2017} find that an accretion disk is likely to be formed in the partial ionization zones in the giant envelope. These zones are located in the outer envelope. Therefore, even if the companion does penetrate somewhat into the envelope, it might still launch jets.  

Numerical simulations of the common envelope evolution that employ no other energy source beside the gravitational energy of the in-spiraling binary system failed to achieve the expected ejection of the common envelope (e.g., \citealt{TaamRicker2010, DeMarcoetal2011, Passyetal2012, RickerTaam2012, Nandezetal2014, Ohlmannetal2016, Staffetal2016MN8, NandezIvanova2016, Kuruwitaetal2016, IvanovaNandez2016, Iaconietal2017b, DeMarcoIzzard2017, Galavizetal2017, Iaconietal2017a}). In light of this results we insert the JFM to the common envelope evolution. 

Another extra energy source {{{ (see some discussion by \citealt{Kruckowetal2016}) }}} that has been suggested to eject the common envelope is the recombination energy of hydrogen and helium (e.g., \citealt{Nandezetal2015, IvanovaNandez2016, {Claytonetal2017}}). However, \cite{Harpaz1998}, \cite{SokerHarpaz2003}, and \cite{Sabachetal2017} question the efficiency by which the recombination energy can be used to eject the envelope.

Along side the negative JFM, where the jets remove envelope mass and hence reduce accretion rate, there is a positive component. The jets remove high entropy gas and energy from the close surroundings of the accreting compact companion, hence reducing the pressure in that region and enabling high accretion rate \citep{Shiberetal2016, Staffetal2016MN}. Without this effect accretion rate would be reduced by the development of high pressure in the close surroundings of accreting compact companion (e.g. \citealt{RickerTaam2012, MacLeodRamirezRuiz2015}). More on the dynamics of jets in the common envelope can be found in the paper by \cite{MorenoMendezetal2017}.  

If jets manage to eject the envelope outside the orbit of the companion  from the onset of the contact between the compact companion and the giant envelope, the system will not enter a common envelope phase. Instead, the system will enter the grazing envelope evolution (GEE; \citealt{SabachSoker2015, Soker2015, Shiberetal2017}). In this case, the system might enter a common envelope evolution only at a later stage. 
The jets might carry more energy than the gravitational energy that is released by the in-spiraling binary system (the core of the giant and the companion). In some cases the orbital separation will not decrease much, or possibly even increase while the jets remove the giant envelope \citep{Soker2017SNIIb}. 

In the present study we conduct our second set of simulations of the GEE. 
In the previous study \citep{Shiberetal2017} we set a constant orbital separation (without spiraling-in), while here we set the system to spiral-in. 
We describe the numerical setting in section \ref{sec:numerics} and the results in section \ref{sec:results}. Our short summary is in section \ref{sec:summary}.

% ==========================================================
\section{NUMERICAL SET-UP}
\label{sec:numerics}
% ==========================================================

We follow the routine as described in \citep{Shiberetal2017}. We run the stellar evolution code \texttt{MESA} (\citealt{Paxtonetal2011}, \citeyear{Paxtonetal2013}, \citeyear{Paxtonetal2015}) and follow the evolution of a star of initial mass $M_{\rm ZAMS}=3.5 M_\odot$ to obtain an asymptotic giant branch (AGB) model with a mass of $M_{\rm AGB} \simeq 3.4 \rmModot$ and radius of $R_{\rm AGB} \simeq 1 \AU$. The star model was imported into the hydrodynamic code \texttt{FLASH} \citep{Fryxell2000}. The grid is a uniform Cartesian three-dimensional cube with side length of $1.2 \times 10^{14} \cm$. {{{{{The size of each cell in the grid is $4.69 \times 10^{11} \cm$ in all runs except the higher resolution run, where the grid has a smaller cell size of $2.34 \times 10^{11} \cm$ }}}}}. We position the stellar giant center at the center of the grid. The gravity of the primary star is taken as constant and spherically symmetric during the simulations. Namely, the influence of mass loss from the envelope and of envelope distortion on gravity is neglected. 

We do not include the gravity of the secondary star. This omission can be partially justified. First, the secondary is a low mass star. Second, the jets interact with the envelope in regions relatively far from the secondary star, where the binding energy is mostly affected by the gravity of the AGB primary star.

We neglect cooling by radiation and treat the gas as an ideal gas with adiabatic index of $\gamma = 5/3$. At the outer layers where the density is lower, radiation can be important and we will address it in our results. To save computational time the inner $0.33 \AU$ sphere of the AGB star is replaced with a constant density and pressure sphere, as we have done in \cite{Shiberetal2017}. In a test run the giant model stays stable, and develops only a weak outflow with a kinetic energy that is negligible compared with the energy of the jets in our simulations. 

We start at $t=0$ by placing a low-mass main sequence secondary star at the surface of the AGB star on its equator, at a $(x,y,z)=(1.5 \times 10^{13} \cm,0,0)$. We do not include, however, the change in the potential of the giant model that would result from the presence of the secondary star. The secondary star is set to orbit counterclockwise at its momentary Keplerian velocity in the plane $z=0$, while performing a spiraling-in motion inside the AGB envelope. During its spiraling-in motion the secondary star launches bipolar jets with a half-opening angle of $\theta_{\rm jet}$ and a velocity of $v_{\rm jet}$. 
 
{{{{{ For the one high-resolution simulation we take $\theta_{\rm jet}=30^\circ$ and $v_{\rm jet}=400 \km \s^{-1}$, about $0.7$ times the escape velocity from the surface of a low mass main sequence star. This is our fiducial run. We also perform lower resolution simulations which one of them is with the same values of $\theta_{\rm jet}=30^\circ,\;v_{\rm jet}=400 \km \s^{-1}$, while in the others we take $\theta_{\rm jet}=60^\circ$ and $v_{\rm jet}=700 \km \s^{-1}$, about $1.2$ times the escape velocity. }}}}}

{{{{{ The jets velocity we use in each run is the terminal velocity of the jets. Namely, this is the velocity at which the jets would have reach infinity without any disturbance.
If we want to include the gravity of the secondary star, we should include also an acceleration mechanism for the jets, or inject the jets with a velocity larger than the terminal velocity. Since we neglect the secondary star gravity we insert the jets with their terminal velocity. 
For example, if we had included the gravity of a solar like secondary star, then we would have injected the jets at $6 R_\odot$ with a velocity of $473 \km \s^{-1}$. }}}}}

When outside the AGB envelope the secondary star accretes mass through an accretion disk (e.g. \citealt{Chenetal2017}) that is likely to launch jets. We assume that when the secondary star enters the envelope it already has an accretion disk that launches jets.    
We let the secondary star to continuously launch two opposite jets from its momentary location. To the initial velocity of the jets relative to the secondary star, $v_{\rm jet}$, we add the momentary orbital velocity of the secondary star (both azimuthal and radial components). {{{{{We numerically inject the jets on the two sides of the secondary star in a cone of a length about the size of two grid cells. Namely, $1\times 10^{12} \cm$ in the low resolution runs and $0.5\times 10^{12} \cm$ in the high resolution run.}}}}}

The inward radial velocity component of the spiraling-in motion is constant and is determined in such a way that in a time of $t_{\rm sp}$ the secondary spirals-in to $2/3$ of its initial radius. In our fiducial run the spiraling-in time $t_{\rm sp}$ was set to $595 \days$, equal to 3 Keplerian orbits around the AGB surface, but due to the spiraling-in motion the secondary completed almost four rounds. {{{{This inward motion was derived based on the spiraling-in orbits that were obtained in former CEE numerical simulations (e.g. \citealt{RickerTaam2012, MacLeodRamirezRuiz2015}). }}}} We also perform two additional runs, one in which we double the spiraling-in time and one where we took half the spiraling-in time. 

The mass injection rate into the two jets in the simulations is 
$ \dot{M}_{\rm jet} =  10^{-3}  \msyr$. This rate is based on Bondi-Hoyle-Lyttleton (BHL) accretion rate and was discussed in the previous paper \citep{Shiberetal2017}. {{{{{ A suppression of the accretion rate relative to the BHL value is expected in these cases due to the pressure built around the accreting object and due to asymmetric accretion flows. However, the suppression is moderate because the jets carry gas and energy out and reduce the pressure \citep{Shiberetal2016}. }}}}}

{{{{{In calculating the BHL accretion rate we consider the giant steep density profile around the companion and not only the local surface density. The BHL accretion rate is then divided by two factors to obtain the mass injection rate into the jets. The first reduction factor of value of ten is the suppression that was discussed above due to pressure around the secondary star. The second reduction factor is a typical ratio of jets mass outflow rate to the accretion rate, which we also take to be ten. On the other hand, we expect an enhancement of about a factor of 15 in the accretion rate as the secondary star spirals-in to the inner denser regions of the envelope. However, we take a conservative approach and keep the mass injection rate at its initial low value of $ \dot{M}_{\rm jet} =  10^{-3}  \msyr$.}}}}} In a forthcoming study the secondary gravity will be included and we will deal with the modulation of the accretion rate as the secondary star dives into the envelope.
 
% ==========================================================
\section{RESULTS}
\label{sec:results}
% ==========================================================

We set the secondary star to spiral-in, and explore the distortion of the envelope and the outflow.
In section \ref{subsec:spiral} we concentrate on the distortion of the envelope, 
and in section \ref{subsec:outflow} we describe the outflow. 
{{{{{In these two subsections we describe a numerical run with one set of parameters, the high-resolution fiducial run and we compare it to a lower resolution run for a resolution study.}}}}} In section \ref{subsec:others} we will present the effects of varying the opening angle of the jets and their velocity, as well as of the spiraling-in time.
 
% ==========================
\subsection{Spiraling-in}
\label{subsec:spiral}
% ==========================

\begin{figure*}
\centering
\includegraphics[trim= 4cm 0.4cm 0.2cm 0.3cm,clip=true,height=0.9\textheight]{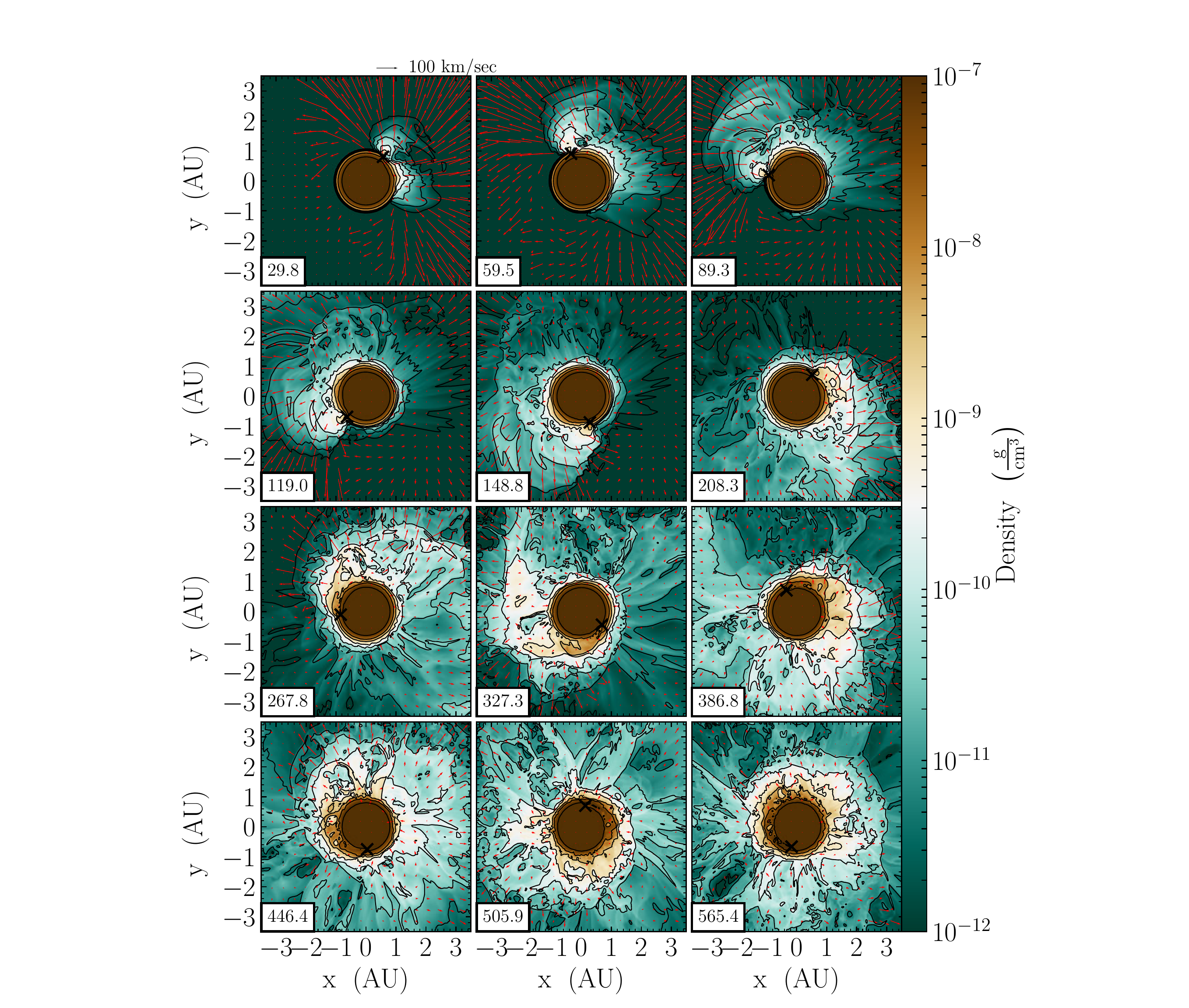} %l b r t
\caption{Density maps and velocity vectors in the equatorial plane $z=0$ of our fiducial run, the only one with the high resolution, with jets' velocity of $v_{\rm jet}=400 \km \s^{-1}$, and a jets' half opening angle of $\theta_{\rm jet}=30^{\circ}$, at twelve different times given in days. The companion orbits counterclockwise, and its momentary location is marked with 'X'. Its initial and final orbital radii are $ 1\AU$ and $0.67 \AU$, respectively. The orbital period at the surface is $198.4 \days$. The length of the arrow is proportional to the gas speed, with scaling of $100 \km \s^{-1}$ as indicated with the arrow above the first panel. We can clearly see how the jets leave a low density trailing behind the secondary star 
as they lift and remove envelope mass.
}
\label{fig:density12}
\end{figure*}

 In Fig. \ref{fig:density12} we present the density of the fiducial run (see section \ref{sec:numerics}) in the equatorial plane at twelve times, as indicated in each panel in days. The location of the companion is marked by an `X', and it orbits counterclockwise. 
 We note the following features. As can be best seen at late times, the secondary penetrates into the envelope, and expels the gas outside its orbit. The expelled gas is seen as envelope inflation trailing the companion. Note that the gravity of the companion is not included, and the effect is solely due to the jets that are launched by the companion. As can be seen by following the low density contours, the outer envelope is heavily distorted and flows outward. By removing the gas, the companion continues to graze the envelope rather than entering a common envelope evolution. By the time we terminate the simulation there is an indication that the companion enters a common envelope, as there is a dense gas outside its orbit. But by this stage the mass accretion rate is likely to increase and so is the energy of the jets. In addition, the release of gravitational energy by the in-spiral orbit should be included by that time. That energy will inflate the envelope and ease its removal by jets. These effects are the tasks of a forthcoming paper.      
% FFFFFFFFFFFFFFFFFFFFFFFFFFFFFFFFFFFFFFFFFFFFFFFFFFFFFF
% FFFFFFFFFFFFFFFFFFFFFFFFFFFFFFFFFFFFFFFFFFFFFFFFFFFFFF

\begin{figure*}
\centering
\includegraphics[trim= 4cm 0.4cm 0.2cm 0.3cm,clip=true,height=0.9\textheight]{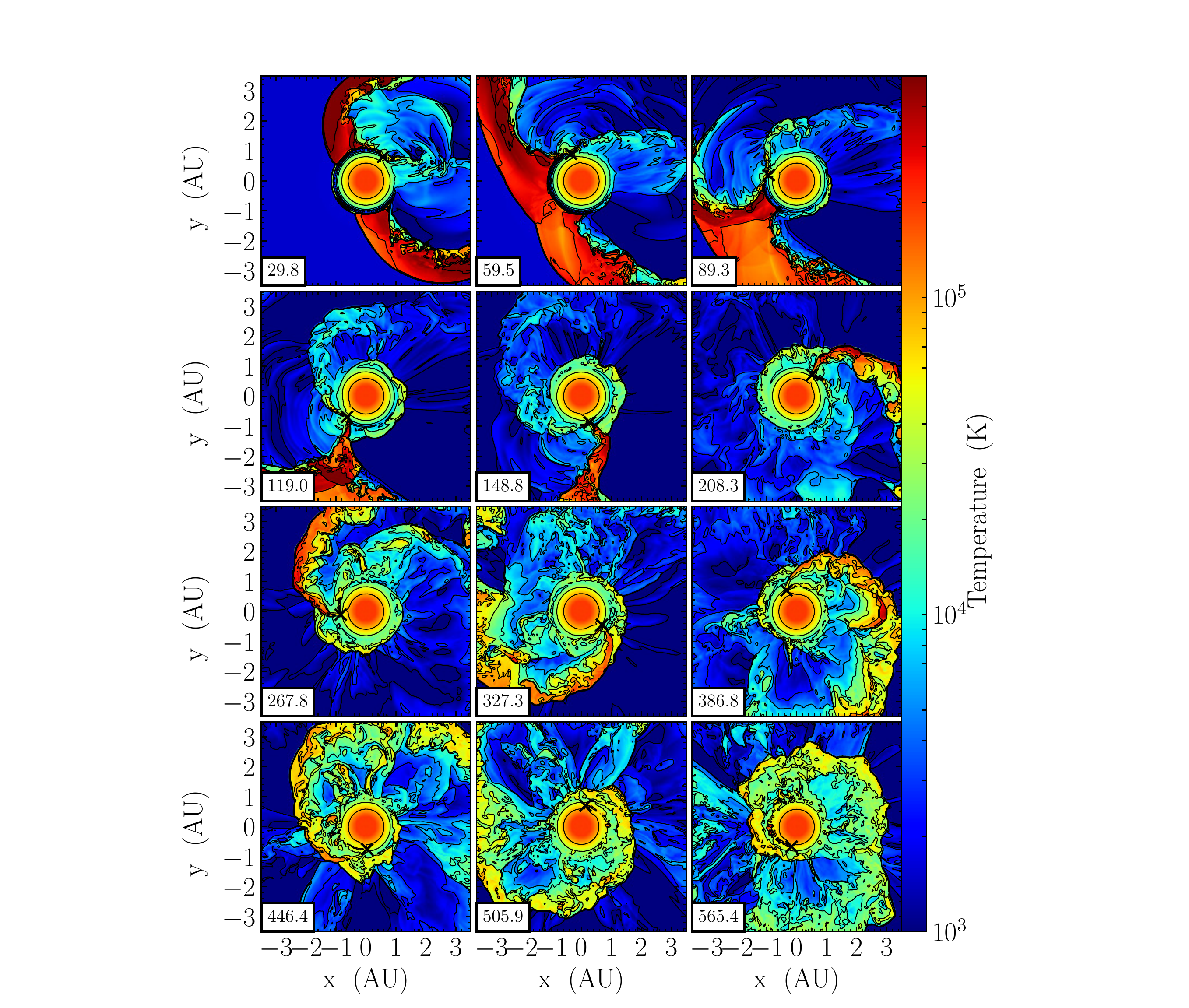} %l b r t
\caption{Like Fig. \ref{fig:density12} but presenting the temperature maps. {{{{ The two red arcs moving around the star in two opposite directions are shocks that are formed by the sudden injection of the jets. They are propagating in very low-density regions, and do not affect much the results. Their signature completely disappear after about 200 days. One should only treat them as a numerical feature.}}}} After this initial phase a spiral structure trailing the secondary star is clearly seen. 
}
\label{fig:temperature12}
\end{figure*}

In fig. \ref{fig:temperature12} we present the temperature in the equatorial plane at the same times as in Fig. \ref{fig:density12}. {{{{ The two arc-like high temperature regions in red are post-shock regions moving around the star. There is a transitory period when these shock waves cross the grid. They are numerically formed by the sudden injection of the jets, but as they propagate in very low density regions they do not influence much the results. After about 200 days their signature disappear, }}}} and a spiral pattern appears to trail the secondary star. The high temperature gas emits X-ray. However, we expect that the dense wind from the giant will absorb the X-ray. The observational signature will be a transient in the visible and in the infra-red band, i.e., an intermediate luminosity optical transient (ILOT), as expected in the formation of some bipolar planetary nebulae \citep{SokerKashi2012}, in the GEE \citep{Soker2016GEE}, and in the common envelope (e.g., \citealt{RetterMarom2003, Retteretal2006, Tylendaetal2011, Tylendaetal2013, Nandezetal2014, Zhuetal2016, Galavizetal2017}). 
% FFFFFFFFFFFFFFFFFFFFFFFFFFFFFFFFFFFFFFFFFFFFFFFFFFFFFF
% FFFFFFFFFFFFFFFFFFFFFFFFFFFFFFFFFFFFFFFFFFFFFFFFFFFFFF

\begin{figure*}
\centering
\includegraphics[trim= 1cm 0.4cm 0.2cm 0.3cm,clip=true,width=0.9\textwidth]{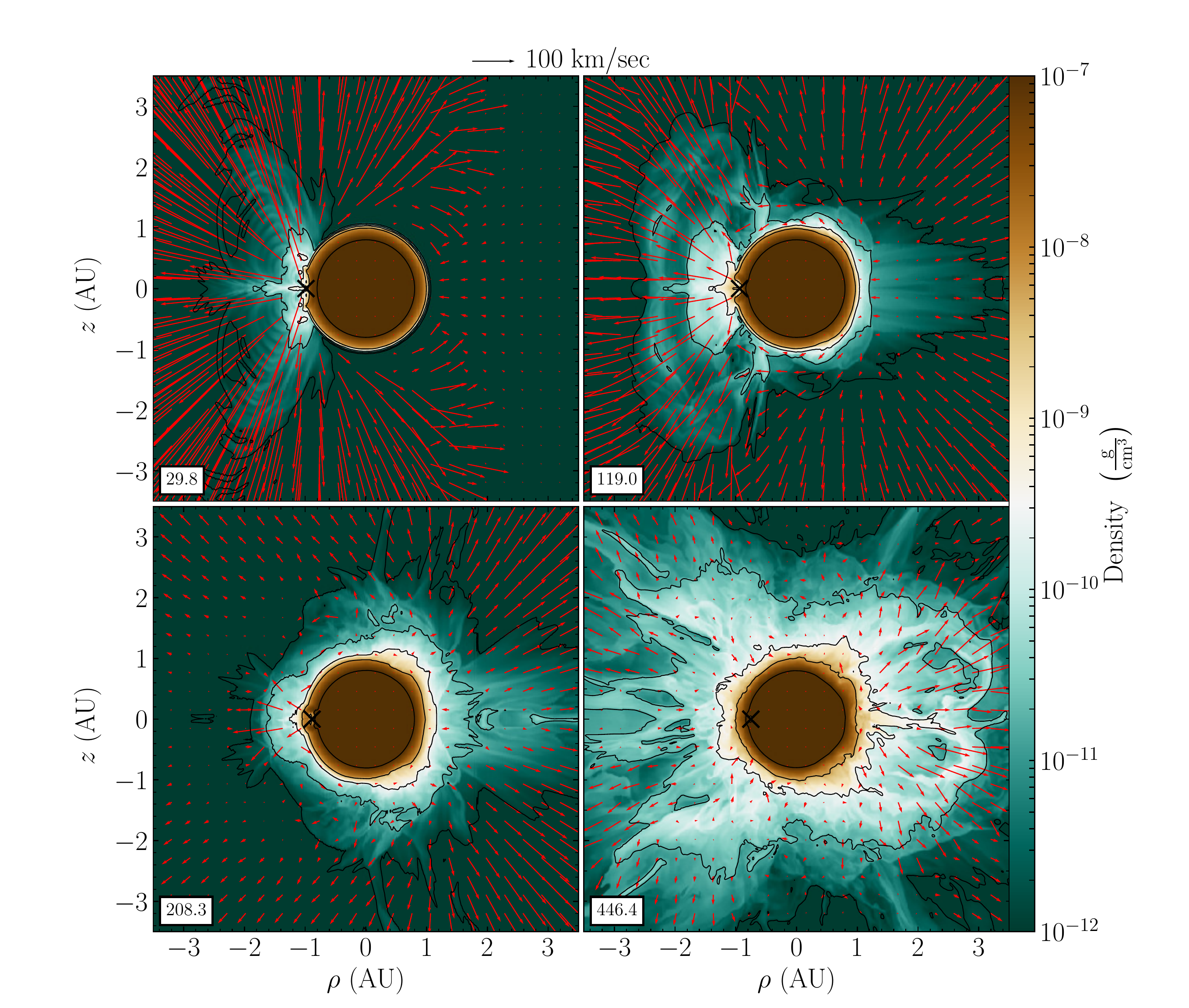} %l b r t
\caption{Density maps and velocity vectors for the fiducial run at four times given in days as indicated, in meridional planes that contains the center of the giant and the momentary location of the secondary star. The horizontal axis is $\rho =\pm \sqrt {x^2+y^2}$. The companion location is marked with 'X'. The different patterns of the flow as the secondary moves inward is clearly seen. The jets proceed out easily when the secondary is in the low density outskirts of the giant, while the jets are halt and bend when the secondary is in the more dens inner regions.
}
\label{fig:4panel_center}
\end{figure*}

To further present the formation of a common envelope at the end of our simulation, in Fig. \ref{fig:4panel_center} we present the flow in meridional planes. 
In the onset of the GEE, when the secondary orbits on the initial giant surface, the jets expel the low density gas at high velocities. As the secondary spirals in, the jets are deflected by the denser inner regions of the giant and bend towards the equatorial plane. The bending of jets inside a common envelope can be seen in earlier simulations \citep{Sokeretal2013, MorenoMendezetal2017}, as well as in the GEE \citep{Shiberetal2017}. 
When the secondary spirals-in even deeper the jets are halt by the giant envelope, and escape only behind the companion in a trailing flow as discussed above. The result is that the secondary enters a common envelope phase. 
Again, the entrance to a common envelope phase might be avoided in full simulations that include the increase in the accretion rate, hence in the power of the jets, and include also the gravitational energy that is released by the in-spiraling companion. This might be the case for relatively more massive companions. We will study the upper limit on the secondary mass for the formation of a common envelope  in the future.    
% FFFFFFFFFFFFFFFFFFFFFFFFFFFFFFFFFFFFFFFFFFFFFFFFFFFFFF
% FFFFFFFFFFFFFFFFFFFFFFFFFFFFFFFFFFFFFFFFFFFFFFFFFFFFFF

\begin{figure*}
\centering
\includegraphics[trim= 1cm 0.2cm 0.2cm 0.3cm,clip=true,width=0.9\textwidth]{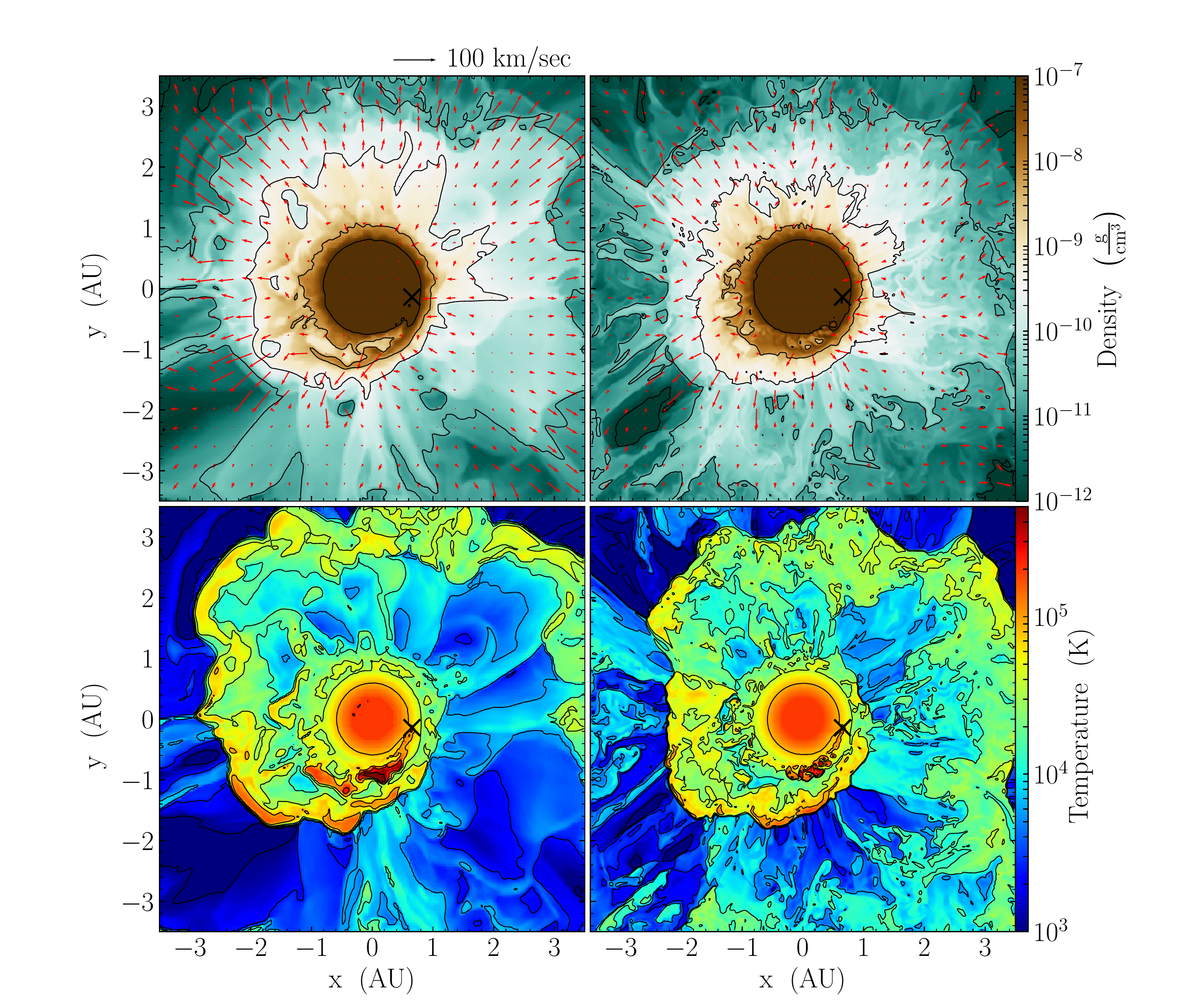} %l b r t
\caption{The hydrodynamic properties in the equatorial (orbital) plane $z=0$ of our fiducial run (right column) and a low resolution run (left column) with identical jet parameters. All panels are at $t=595~$days, amounts to almost 4 orbital rounds, and when the orbital separation has reduced to $0.67\AU$. The upper panels show density and velocity maps, and the lower panels show the temperature maps. 
}
\label{fig:panel_40030_end}
\end{figure*}

\begin{figure*}
\centering
\includegraphics[trim= 1cm 0.2cm 0.2cm 0.3cm,clip=true,width=0.9\textwidth]{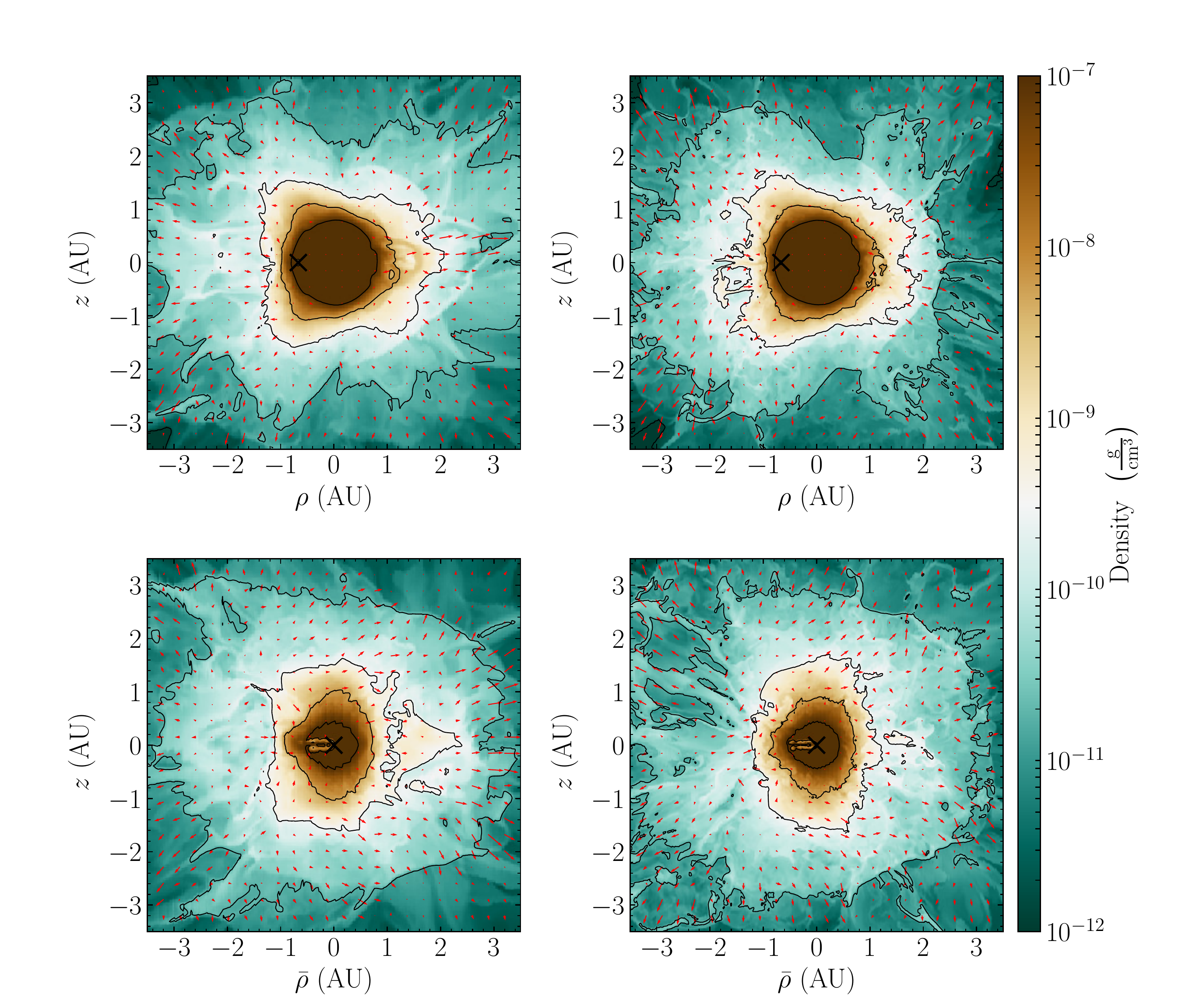} %l b r t
\caption{Like upper panels of Fig. \ref{fig:panel_40030_end}, but for different planes. 
Again, our fiducial run is on the right column and the low resolution run with identical jet parameters is on the left column. The upper row shows the density and velocity in the $\rho-z$ meridional plane that contains the center of the giant and the momentary location of the secondary star (the same plane as depicted in Fig \ref{fig:4panel_center}; $\rho =\pm \sqrt {x^2+y^2}$). The lower row shows the density and velocity in the $\bar{\rho}-z$ plane that is perpendicular to the momentary radius vector of the secondary star and perpendicular to the equatorial plane. $\bar{\rho}=\pm \sqrt {(x-x_s)^2+(y-y_s)^2}$ in these panels, where $(x_s,y_s,0)$ is the momentary three dimensional position of the secondary.
}
\label{fig:panel_40030_end_2}
\end{figure*}

{{{{{Fig. \ref{fig:panel_40030_end} and \ref{fig:panel_40030_end_2} show the flow properties at the end of two simulations after the secondary has completed about 4 rounds.
One run is the fiducial run (right column), and the other run has identical parameters but with a lower resolution of twice the cells size (left column). Fig. \ref{fig:panel_40030_end} presents the flow in the equatorial plane. A spiral arm structure is clearly seen, mainly in the temperature maps (lower panels). 
The flow structure is similar in both resolutions, but the higher resolution run reveals finer details. }}}}}
% FFFFFFFFFFFFFFFFFFFFFFFFFFFFFFFFFFFFFFFFFFFFFFFFFFFFFF
% FFFFFFFFFFFFFFFFFFFFFFFFFFFFFFFFFFFFFFFFFFFFFFFFFFFFFF
% FFFFFFFFFFFFFFFFFFFFFFFFFFFFFFFFFFFFFFFFFFFFFFFFFFFFFF
% FFFFFFFFFFFFFFFFFFFFFFFFFFFFFFFFFFFFFFFFFFFFFFFFFFFFFF

{{{{{In the upper panels of Fig. \ref{fig:panel_40030_end_2} }}}}} we present the density map in a meridional plane as in Fig. \ref{fig:4panel_center}. The outer contours of the density maps show a distorted envelope lifted by the energetic jets. In the lower panel we present the density map in the plane perpendicular to the momentary radius vector of the secondary star (a plane tangential to the $\phi$ direction) and perpendicular to the equatorial plane. In these panels the secondary is at the center. The jets are strongly bent backward, as seen in the two narrow light-brown stripes to the left of the secondary star. Over all, the jets that are launched from the spiraling-in secondary star lead to an asymmetric and complicated flow structure. 
   
To summarized this subsection we emphasize the following. 
(1) The jets expel mass efficiently at the beginning of the simulation when the secondary star orbits the giant outskirts. (2) Deeper in the envelope the jets expel mass less efficiently and the outflow is directed mainly to the equatorial plane in a narrow region trailing the companion. The common envelope evolution then commences. (3) We note though that in our simulations we do not include the gravitational energy that is released by the in-spiraling process. The inclusion of the orbital energy that is released will delay the onset of the common envelope, or will avoid it altogether.
The later is expected for relatively massive companions.

% ==========================
\subsection{The outflow}
\label{subsec:outflow}
% ==========================

\begin{figure}
\includegraphics[trim= 0.0cm 0.6cm 0.0cm 0.4cm,clip=true,width=0.99\columnwidth]{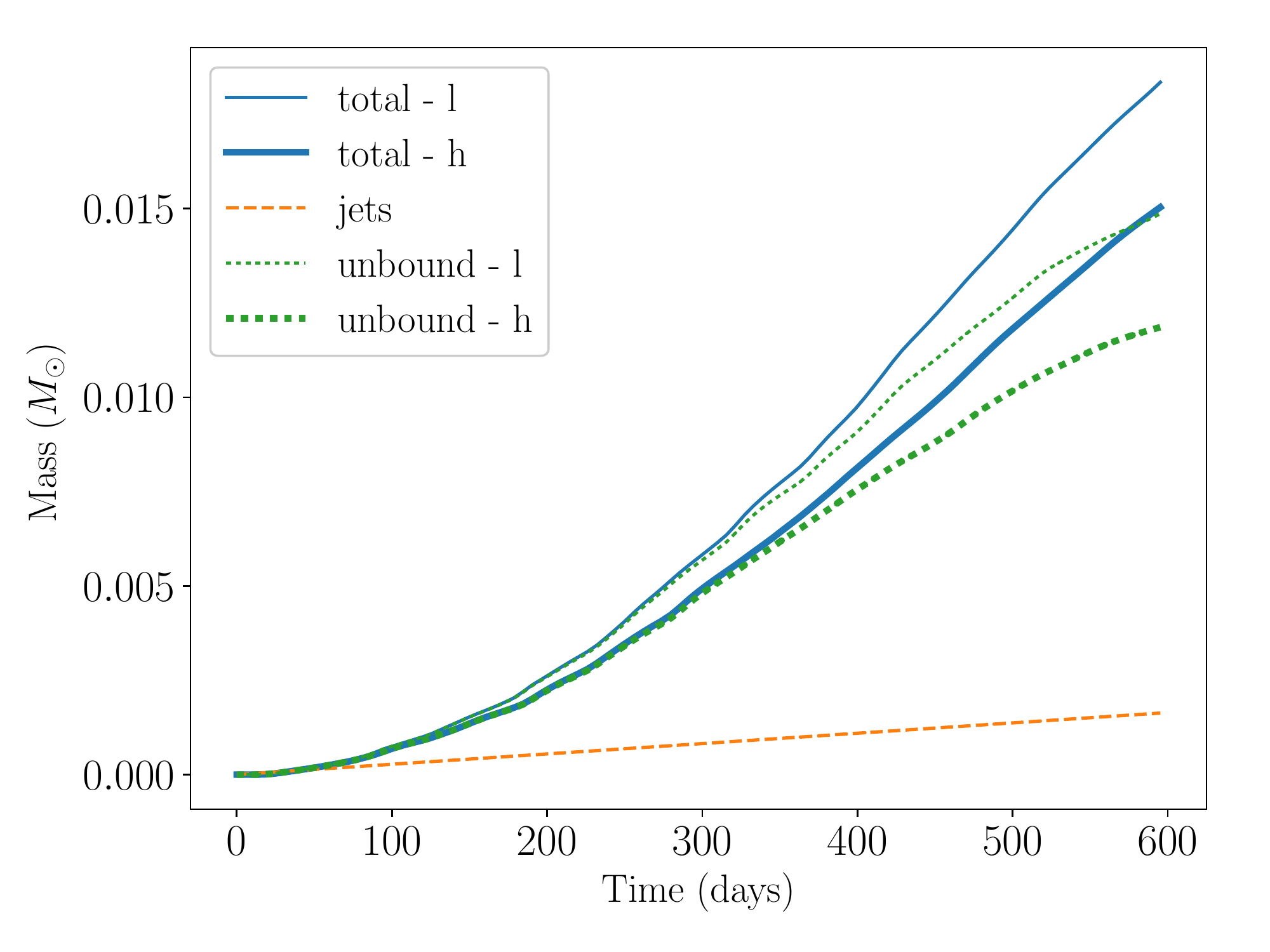}
\caption{The total mass (solid blue line) that crosses out through a spherical shell of radius $4 \AU$, the unbound (dotted green line) mass that crosses out the same sphere, and the injected mass into the jets (dashed orange line) as function of time. The amount of mass flowing outward is more than ten times higher than the mass injected in the jets. The unbound mass percentage of the total mass decreases as the secondary spirals-in. {{{{The thick lines are for the high resolution run while the thin lines are for the low resolution run.}}}} }
\label{fig:Mout}
\end{figure}

We start by presenting some average properties of the mass ejected from the binary system.
In Fig. \ref{fig:Mout} we show the evolution of the total mass that is ejected from the system in blue solid line, and the part that is unbound, i.e., has a positive total energy, in dotted green line. We compare these two quantities to the mass that is injected into the jets (dashed orange line). The amount of ejected mass is calculated as mass leaving a sphere of radius $4 \AU$.
{{{{{The thick lines are for the high resolution fiducial run and the thin lines are for the lower resolution run with identical jet parameters. }}}}}
We emphasize again that the only extra energy that is included is that of the jets, and in this paper we do not yet include the gravitational energy released by the in-spiraling binary system. We note that at late times, when the secondary star dives into the envelope the increase in ejected unbound mass decreases.

{{{{{ We see from Fig. \ref{fig:Mout} that more mass is ejected and becomes unbound in the lower resolution run, by about $16 \%$ at the end of the runs. Although the runs have the identical physical jet parameters, i.e., the jet velocity and the jet opening angle, they differ in their numeric parameters. Mainly, the cone length in which the jets are numerically injected is different. In the high resolution run the cone length is half of the cone length in the low resolution. Although the jets have the same injection rate in both cases, in the high resolution run the jets are injected inside a smaller volume. The low resolution run where the jets are numerically injected inside a larger volume yields higher mass ejection. }}}}}
% FFFFFFFFFFFFFFFFFFFFFFFFFFFFFFFFFFFFFFFFFFFFFFFFFFFFFF
% FFFFFFFFFFFFFFFFFFFFFFFFFFFFFFFFFFFFFFFFFFFFFFFFFFFFFF

\begin{figure}
\includegraphics[trim= 0.0cm 0.6cm 0.0cm 0.4cm,clip=true,width=0.99\columnwidth]{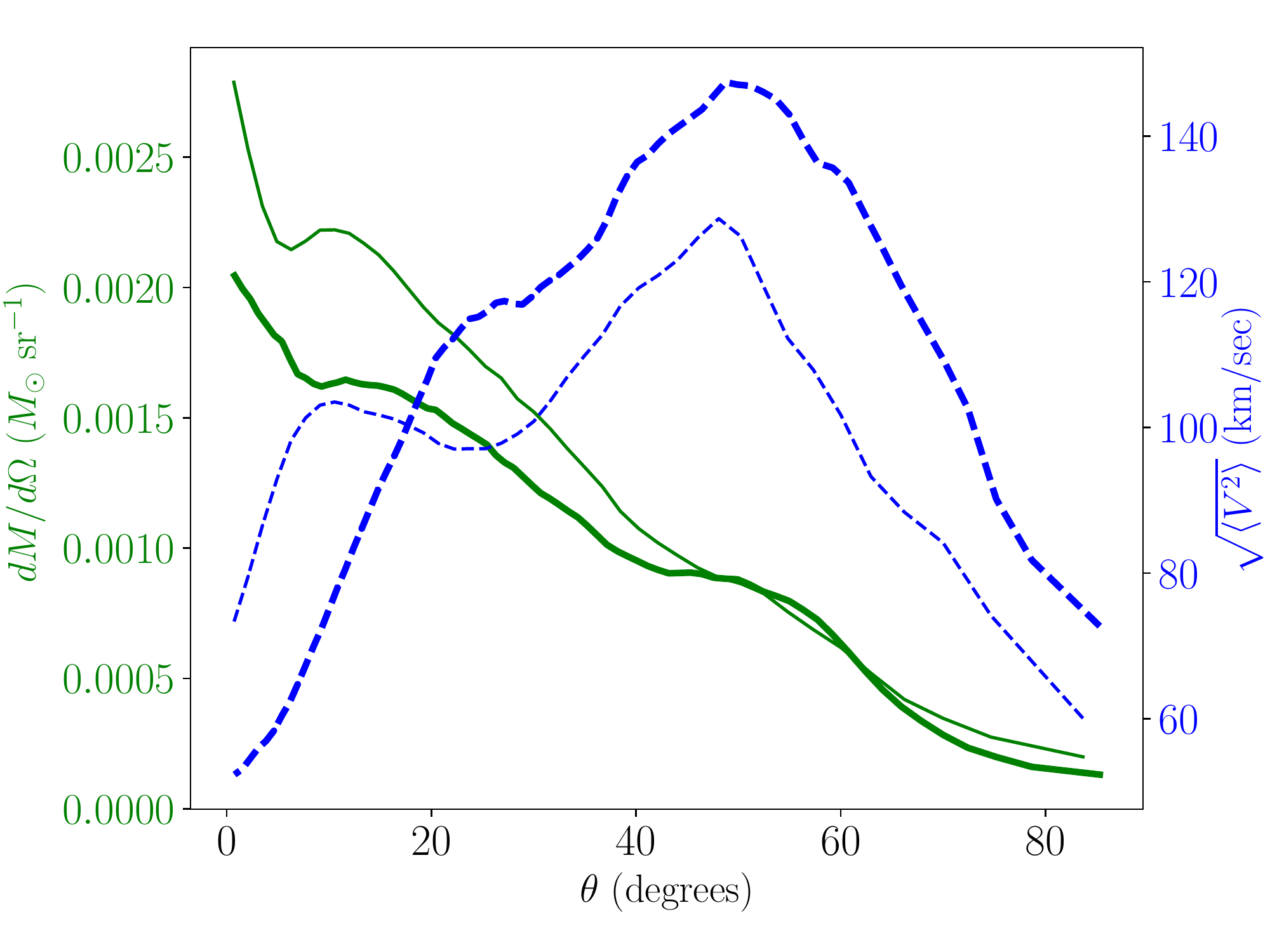}
\caption{ The total mass crosses out through a spherical shell of radius $4 \AU$ per unit solid angle (green line) and the root mean square of the outflow velocities that crossed the same sphere (dashed blue line) as a function of the angle from the equatorial plane $\theta$,
after $595 \days$ of the fiducial Run.
The equator is at $\theta=0^\circ$ and the poles are at $\theta=90^\circ$. The amount of mass lost is that from the two hemispheres combined. Although the mass loss is concentrated near the equatorial plane, the highest average outflow velocity is at mid-latitude. {{{{The thick lines are for the high resolution fiducial run while the thin lines are for the low resolution run with identical jet parameters.}}}} }
\label{fig:mass_angle}
\end{figure}

In Fig. \ref{fig:mass_angle} we present the mass per unit solid angle, $dm/d\Omega$, that left a sphere of radius $4 \AU$ over 595 days (about 4 orbits), as a function of the angle from the equatorial ($\theta=0^\circ$ is the equatorial plane and $\theta =90^\circ$ is the poles). The quantity $dm/d\Omega$ is calculated by summing over the azimuthal angle $\phi$ at each angle $\theta$, and combining the two hemispheres.  
We also show the value of the root mean square of the velocity $\sqrt{\langle v^2 \rangle} \equiv \sqrt{ 2 dE_k(\theta) / dM(\theta)}$, where $d E_k(\theta)$ and  $d M(\theta)$ are the kinetic energy and mass, respectively, that left the $4 \AU$ sphere. {{{{{ The thick lines show results for the fiducial run while the thin lines are for the low resolution run with identical jets parameters. }}}}} The outflow is concentrated around the equatorial plane in both cases, although it is not monotonic with $\theta$. What we find more interesting, is that the maximum outflow velocity is at mid-latitude (see also \citealt{Shiberetal2017}). {{{{{In the high resolution run the maximal velocity is higher with a more pronounced pick. The outflow is faster in this case but less massive.}}}}} 
% FFFFFFFFFFFFFFFFFFFFFFFFFFFFFFFFFFFFFFFFFFFFFFFFFFFFFF
% FFFFFFFFFFFFFFFFFFFFFFFFFFFFFFFFFFFFFFFFFFFFFFFFFFFFFF

\begin{figure}
\includegraphics[trim= 0.0cm 0.6cm 0.0cm 0.4cm,clip=true,width=0.99\columnwidth]{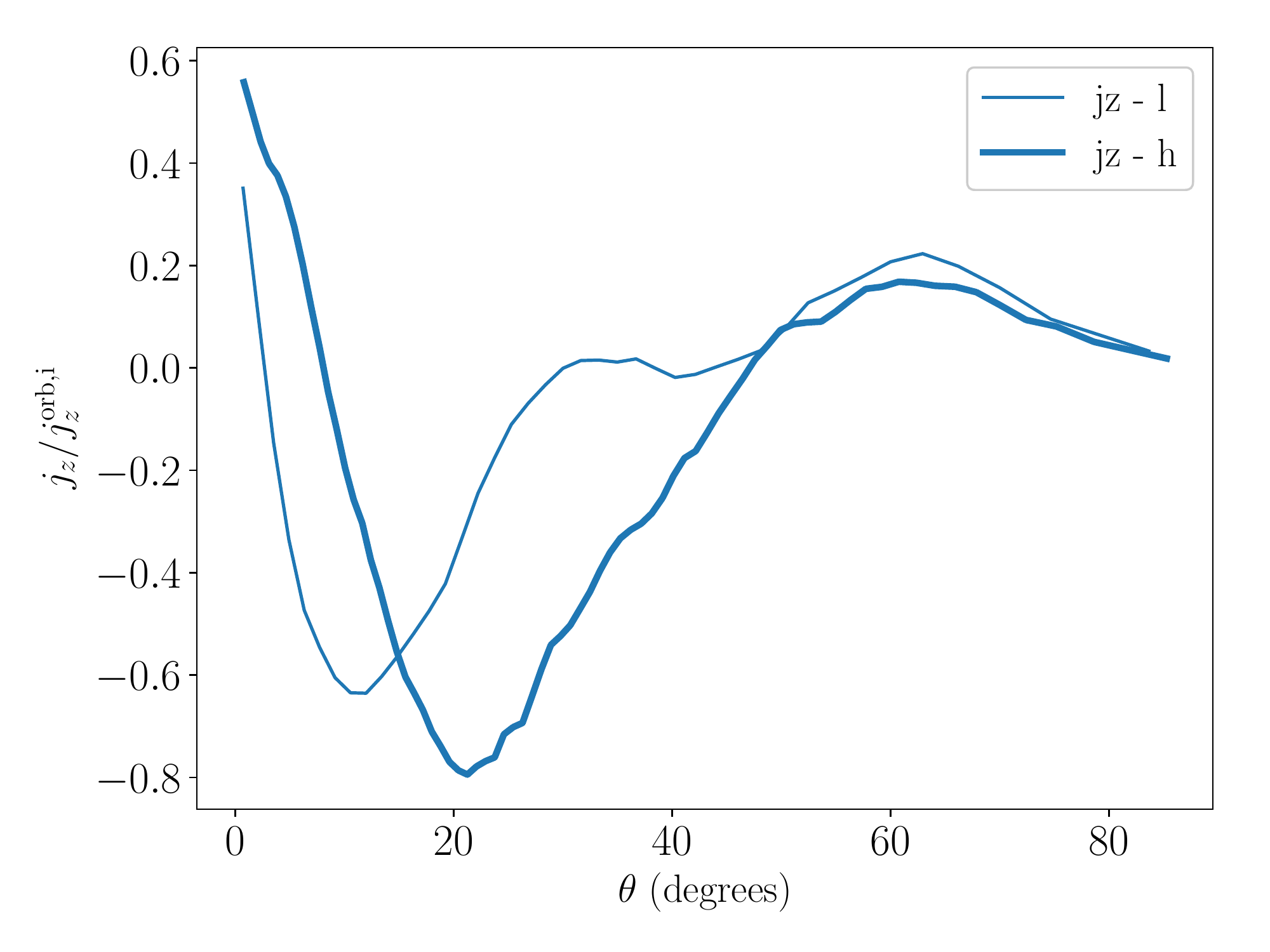}
\caption{The $z$ component of the specific angular momentum of the ejected mass, as function of angle from the equatorial plane $\theta$, and relative to the specific angular momentum of a body performing Keplerian motion on the initial surface of the AGB star. The thick line is from the high resolution run while the thin line is from the low resolution run.}
\label{fig:angle_angular}
\end{figure}

In Fig. \ref{fig:angle_angular}, we show the specific angular momentum of the outflowing gas about the axis perpendicular to the orbital plane $j_z$ in the fiducial run (thick line) and in the low resolution run (thin line). This is similar to the mass per unit solid angle and the velocity of the outflowing gas as function of $\theta$ that we presented in Fig. \ref{fig:mass_angle}, but for specific angular momentum.   
We show the specific angular momentum of the outflowing gas relative to the specific angular momentum of a body performing Keplerian motion on the initial surface of the AGB star $j^{{\rm orb},i}_{z}=\sqrt{GM_{{\rm AGB}}R_{AGB,i}}$. We find the total average specific angular momentum of the ejected mass in {{{{{both cases}}}}} to be $j_z({\rm total}) \simeq - 0.2 j^{{\rm orb},i}$. The negative values imply that the ejected mass carries angular momentum opposite to that of the secondary star. This is because the jets are bent backward and push gas in the opposite direction. The jets can actually spin-up the envelope, but not by much. 
% FFFFFFFFFFFFFFFFFFFFFFFFFFFFFFFFFFFFFFFFFFFFFFFFFFFFFF
% FFFFFFFFFFFFFFFFFFFFFFFFFFFFFFFFFFFFFFFFFFFFFFFFFFFFFF

\begin{figure*}
\centering
\includegraphics[trim= 1cm 0.4cm 0.2cm 0.3cm,clip=true,width=0.45\textwidth]{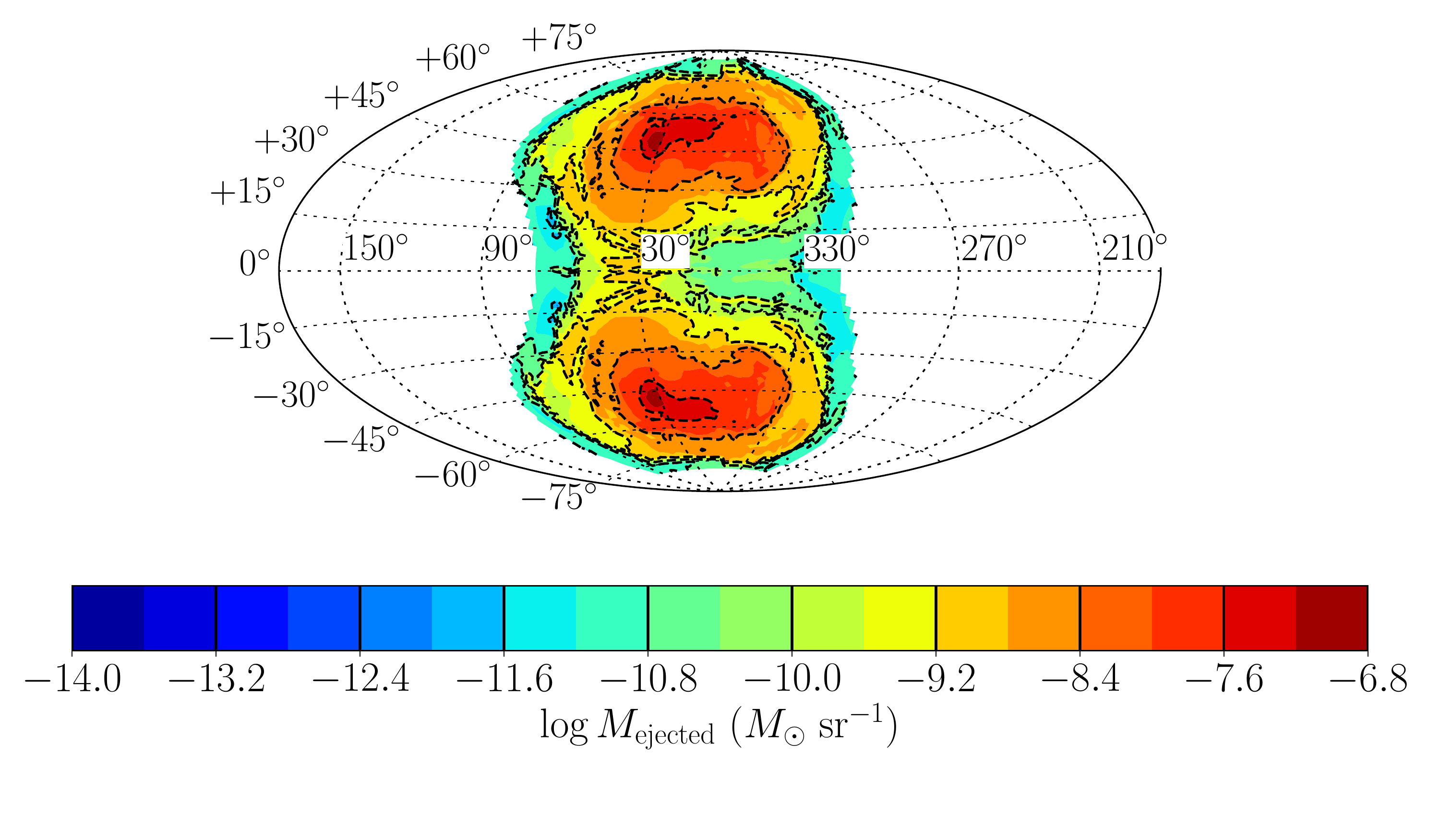}
\includegraphics[trim= 1cm 0.4cm 0.2cm 0.3cm,clip=true,width=0.45\textwidth]{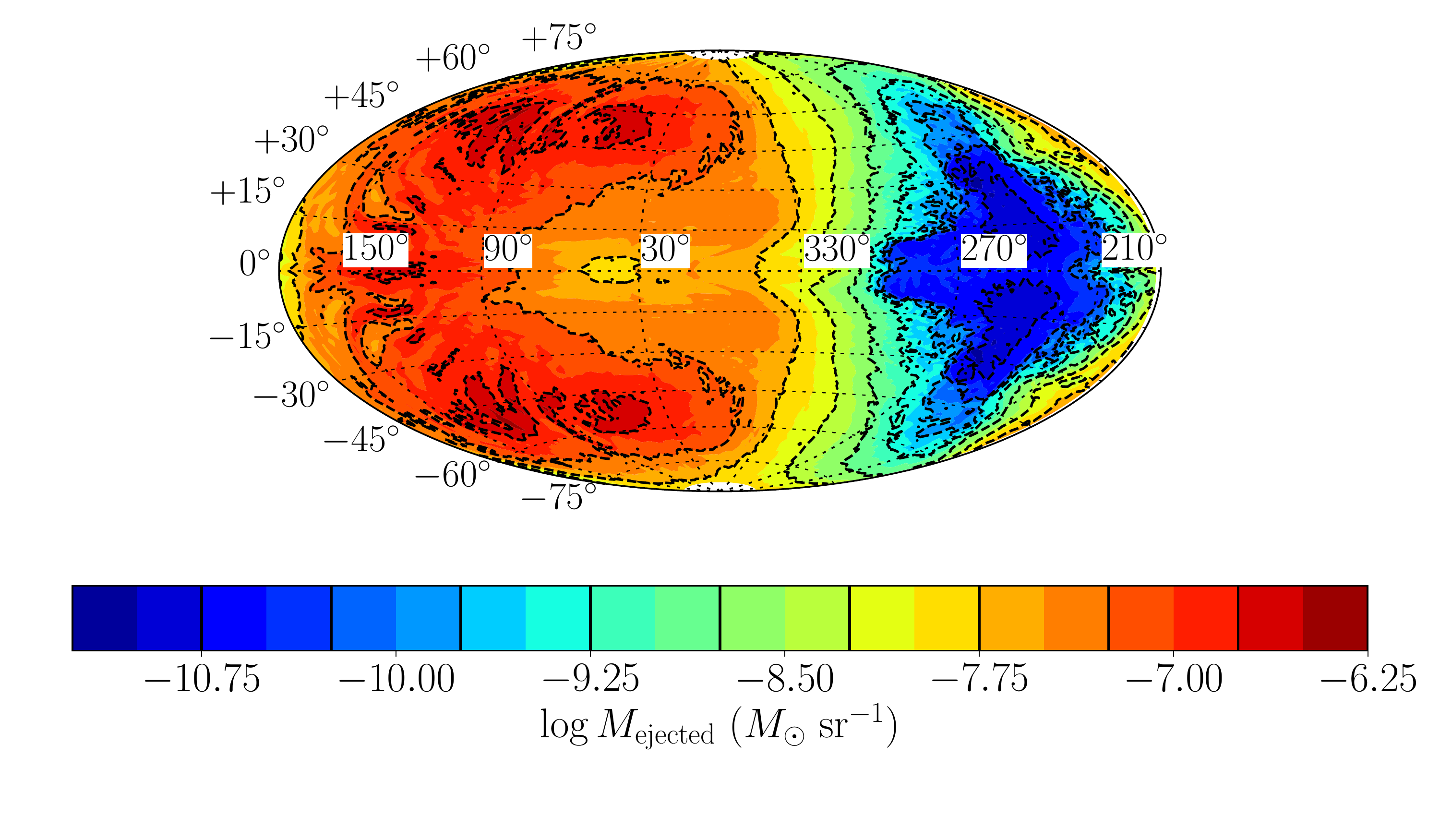}
\hspace*{0.4cm}
\vspace*{0.4cm}
\includegraphics[trim= 1cm 0.4cm 0.2cm 0.3cm,clip=true,width=0.45\textwidth]{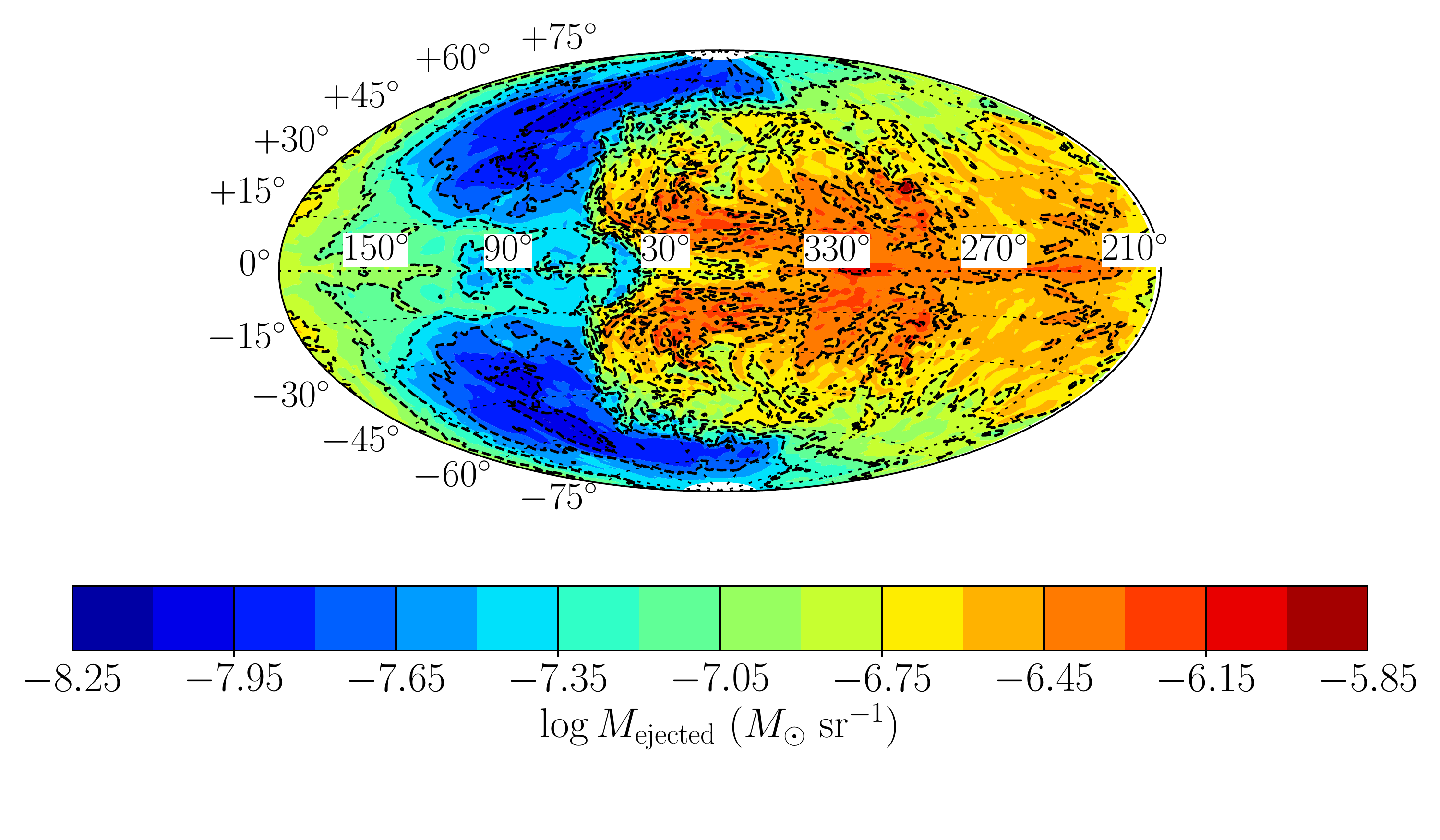}
\includegraphics[trim= 1cm 0.4cm 0.2cm 0.3cm,clip=true,width=0.45\textwidth]{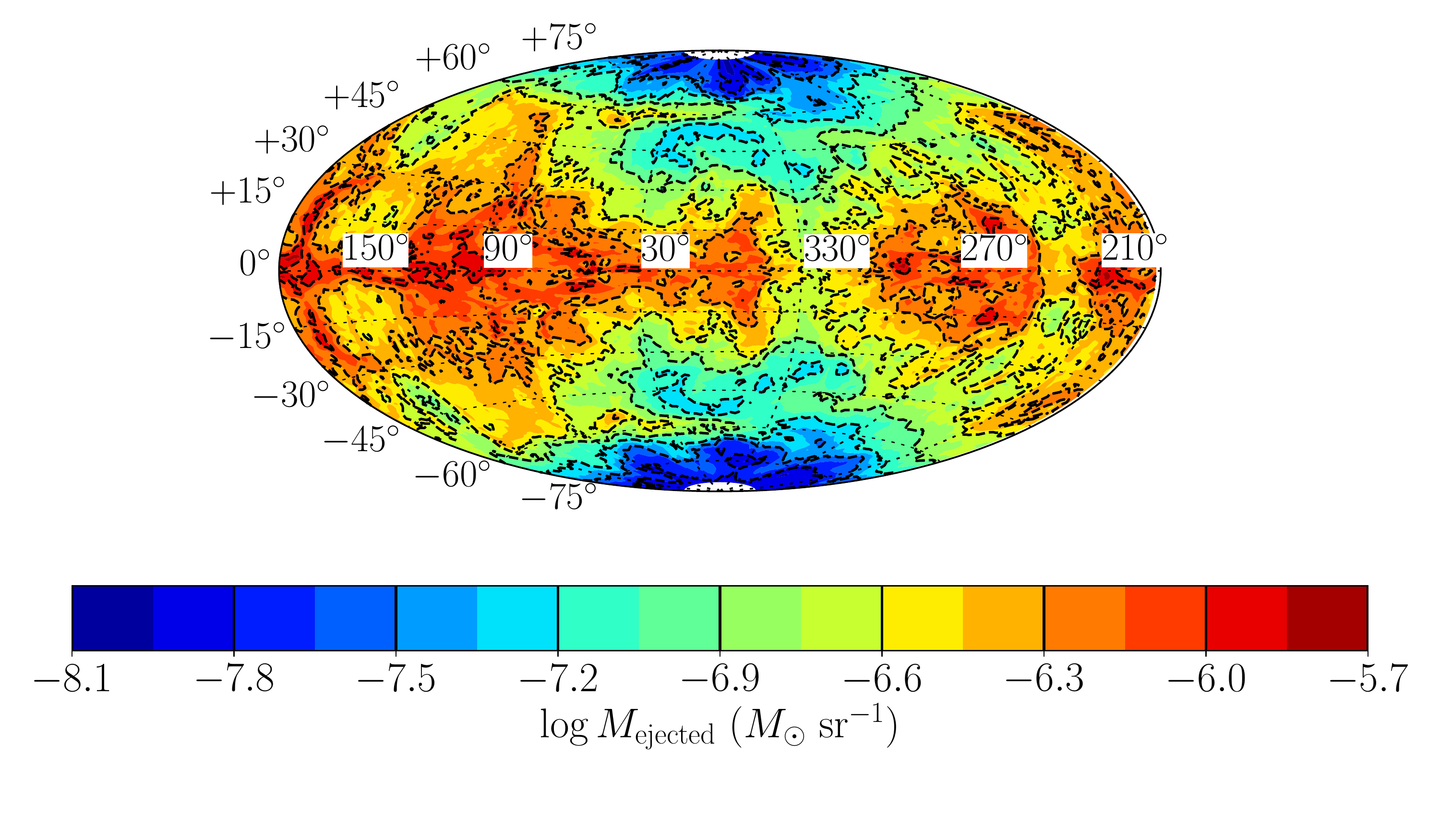}
\hspace*{0.4cm}
\includegraphics[trim= 1cm 0.4cm 0.2cm 0.3cm,clip=true,width=0.7\textwidth]{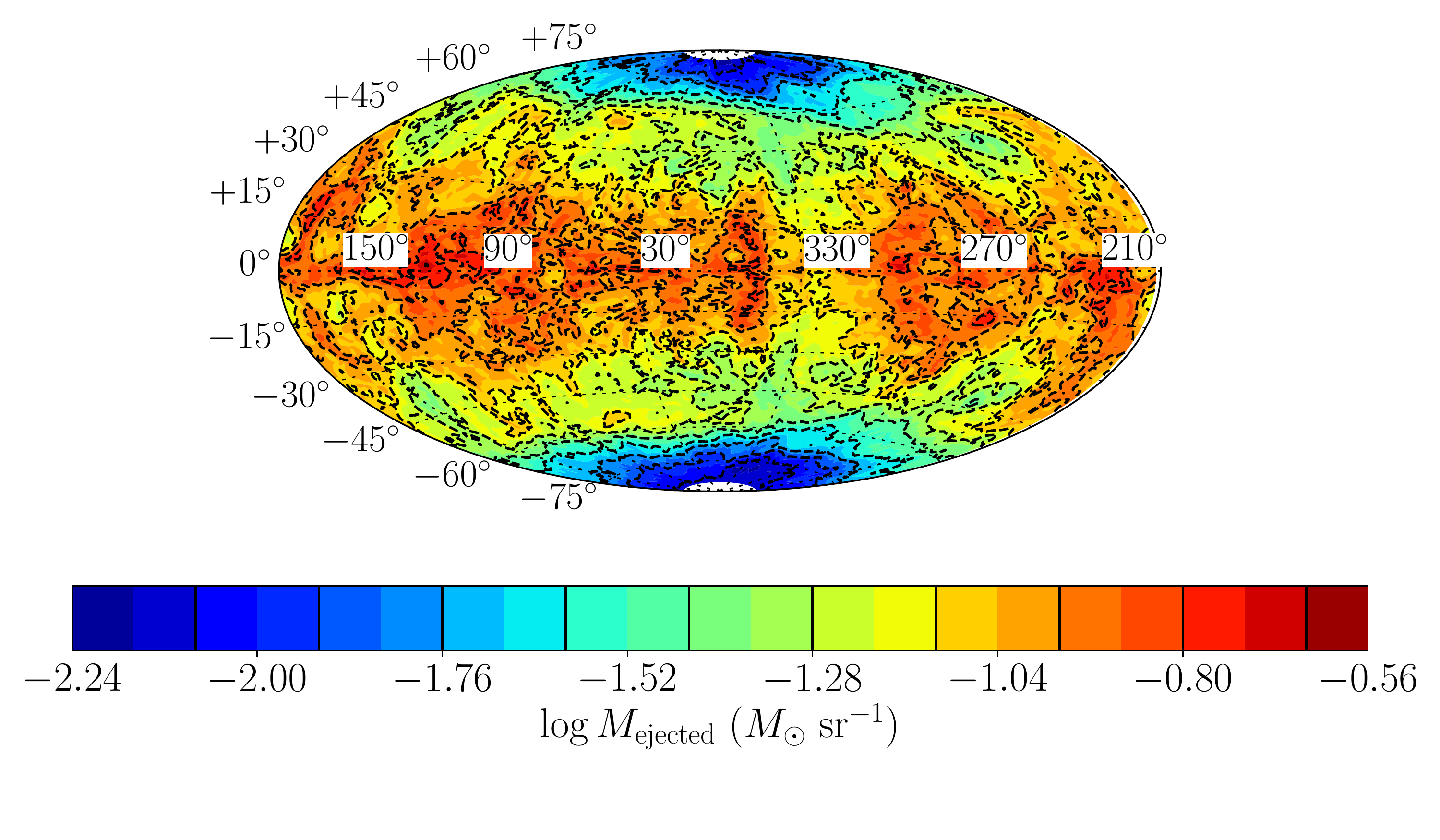}
\caption{The total mass outflow from a spherical shell of radius $4 \AU$ per unit solid angle during five time periods. The four smaller maps correspond to the time periods of $0-29.8$, $29.8-119$, $119-267.9$, and $267.9-446.4 \days$. The lower bigger map corresponds to the total mass outflow per unit solid angle over the entire simulation, $0-595 \days$. Zero latitude is the equatorial plane and zero longitude (at the center) is the initial location of the companion. The companion is moving towards higher angles, namely from the right to the left. We can see a concentration of clumpy mass ejection near the equatorial.
}
\label{fig:outflow_map}
\end{figure*}

In Fig \ref{fig:outflow_map} we present the ejected mass per unit solid angle over different times of the {{{{{fiducial run, }}}}} as function of $(\theta, \phi)$ as Hammer projection of a sphere of radius $4 \AU$. 
The evolution of the ejected mass geometry clearly shows the highly asymmetrical mass ejection and its concentration near the equatorial plane. These maps also show the pronounced clumpy nature of the outflow. At later times the concentration of outflow toward the equatorial plane increases. This results from a stronger bending backward of the jets as the secondary dives deeper into the envelope. The location of the strongest outflow moves around because the ejection has a spiral morphology (see Figs. \ref{fig:density12} and \ref{fig:temperature12}). 
% FFFFFFFFFFFFFFFFFFFFFFFFFFFFFFFFFFFFFFFFFFFFFFFFFFFFFF
% FFFFFFFFFFFFFFFFFFFFFFFFFFFFFFFFFFFFFFFFFFFFFFFFFFFFFFF

\begin{figure*}
\centering
\includegraphics[trim= 1cm 0.4cm 0.2cm 0.3cm,clip=true,width=0.45\textwidth]{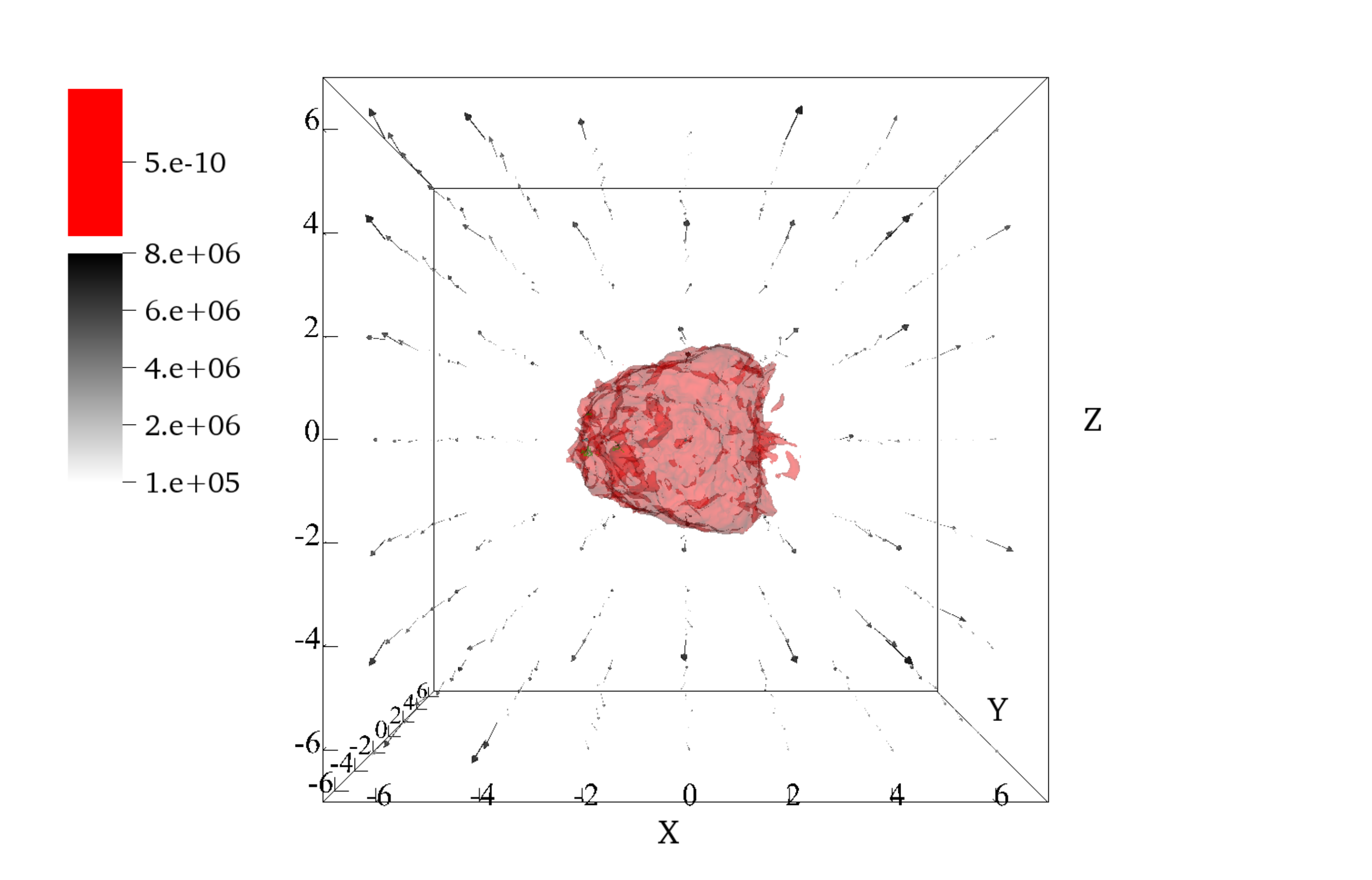}
\includegraphics[trim= 1cm 0.4cm 0.2cm 0.3cm,clip=true,width=0.45\textwidth]{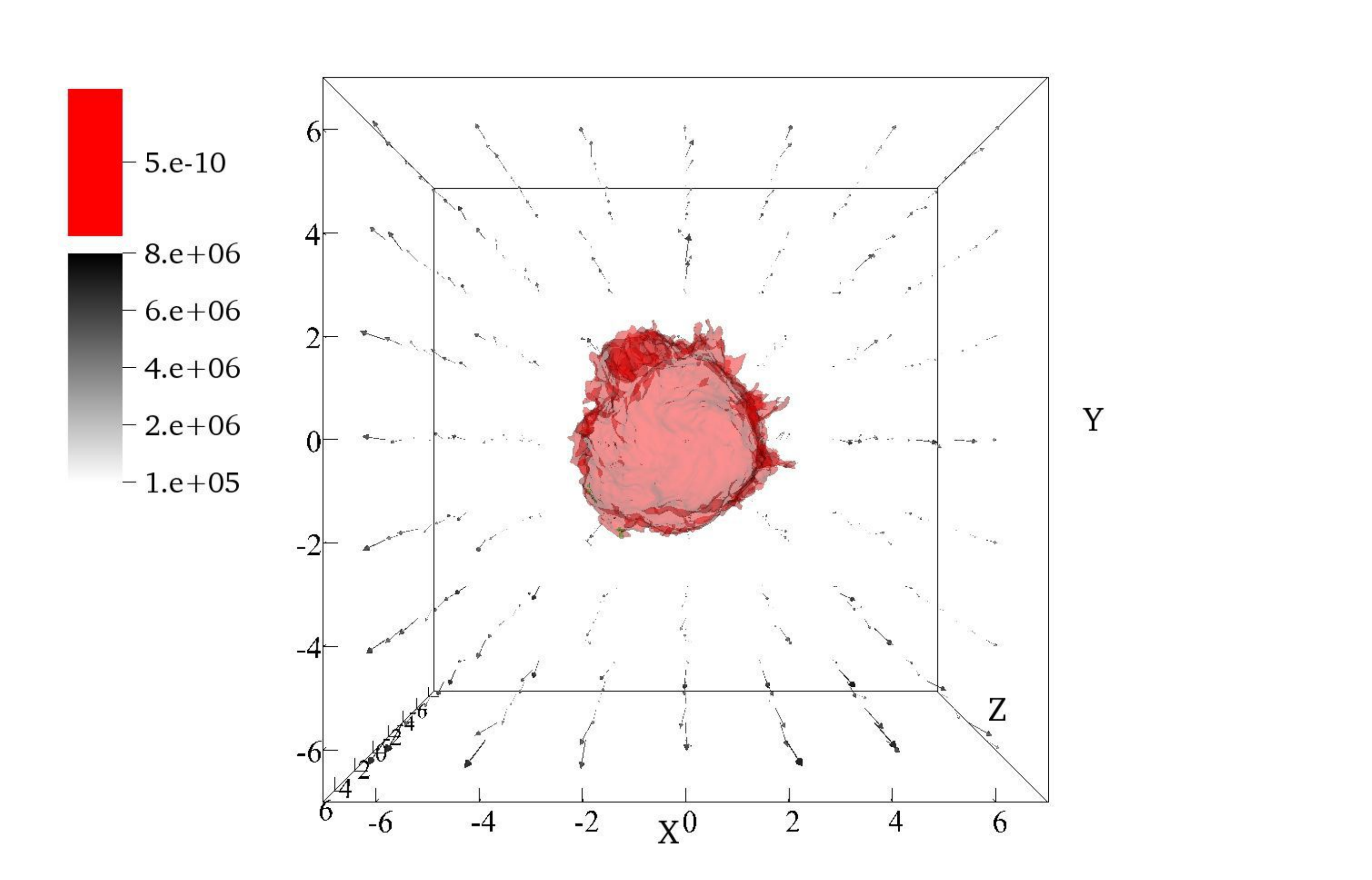}
%\hspace*{0.4cm}
%\vspace*{0.4cm}
\includegraphics[trim= 1cm 0.4cm 0.2cm 0.3cm,clip=true,width=0.45\textwidth]{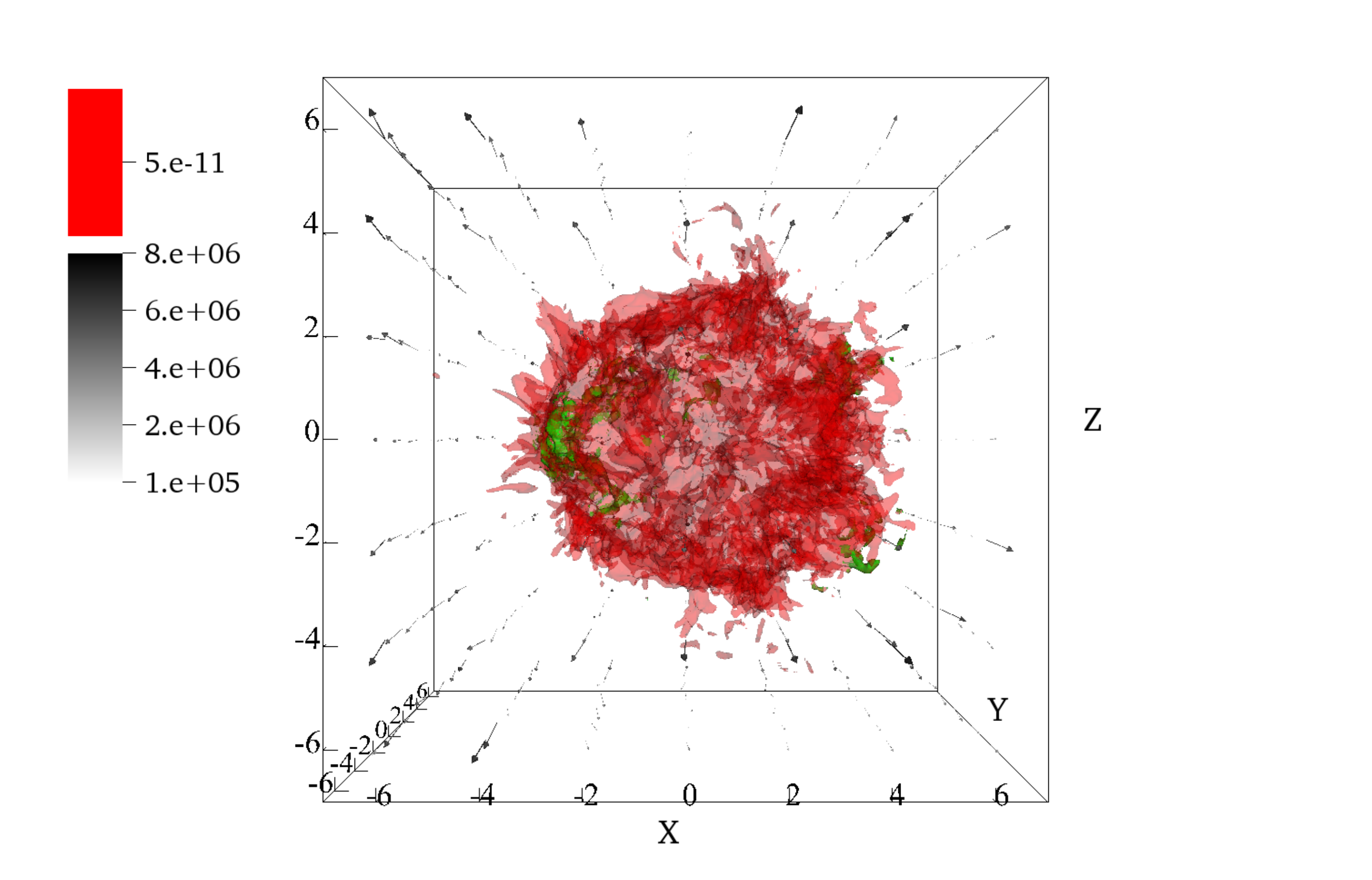}
\includegraphics[trim= 1cm 0.4cm 0.2cm 0.3cm,clip=true,width=0.45\textwidth]{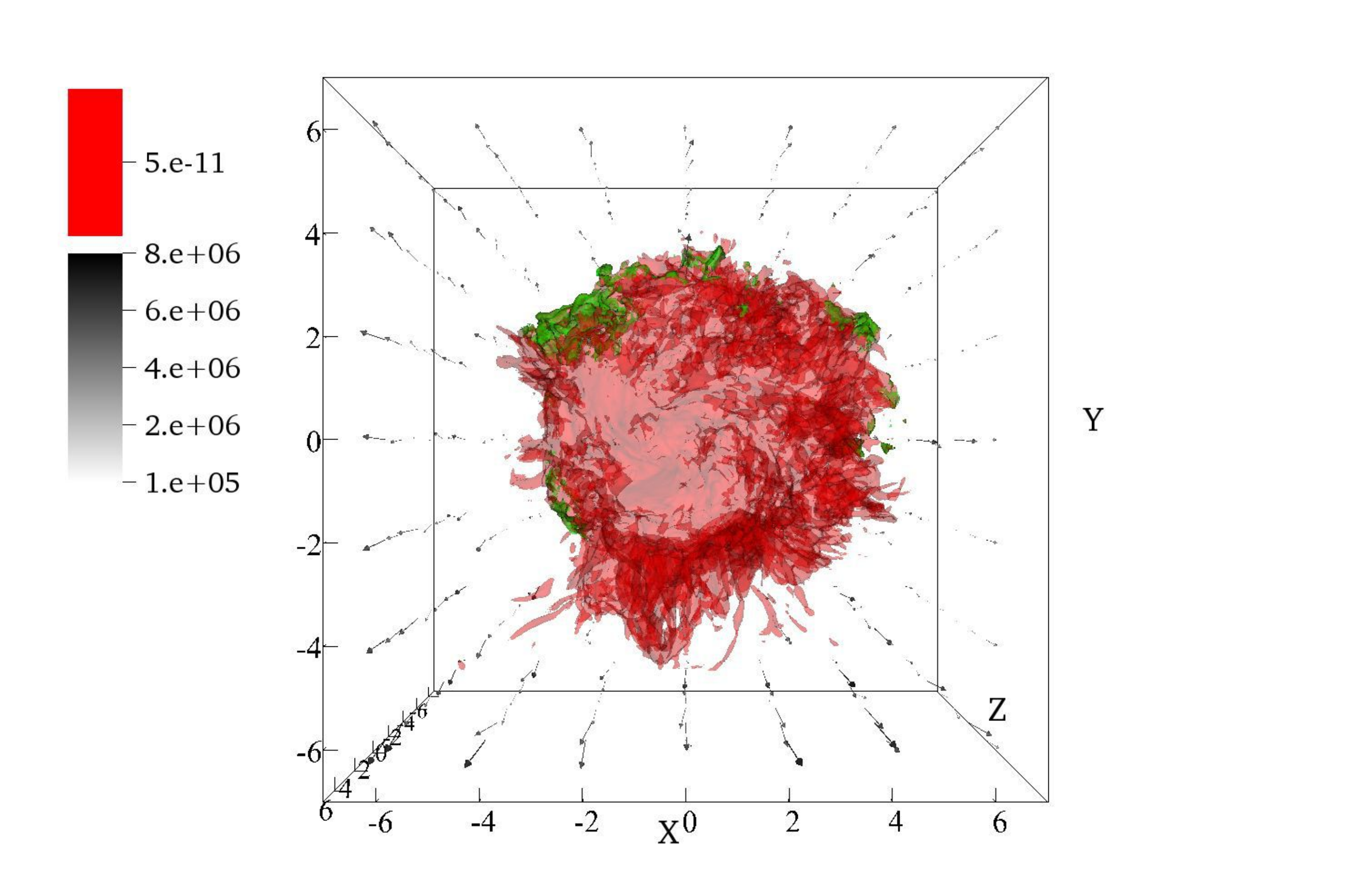}
%\hspace*{0.4cm}
\includegraphics[trim= 1cm 0.4cm 0.2cm 0.3cm,clip=true,width=0.45\textwidth]{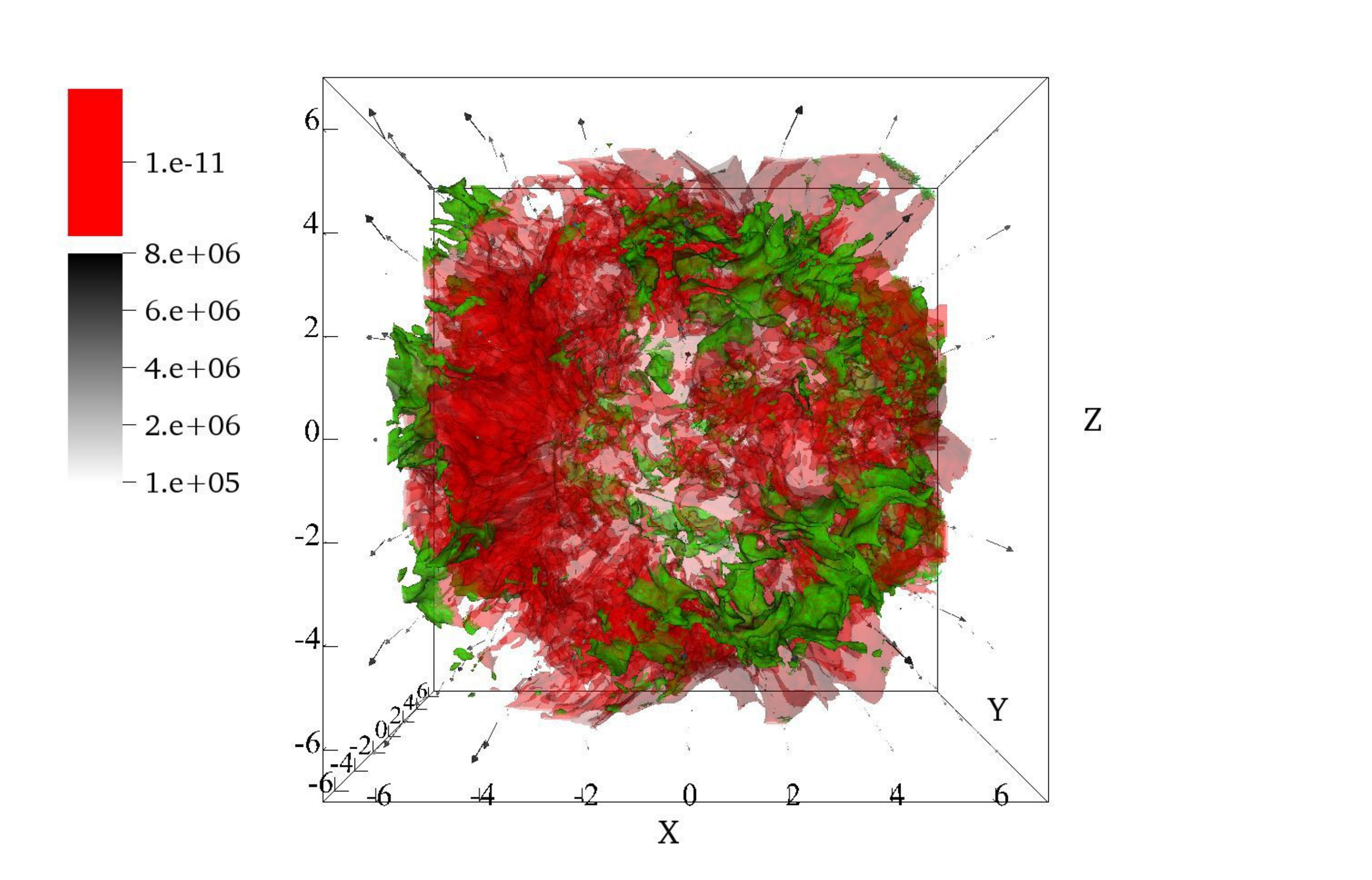}
\includegraphics[trim= 1cm 0.4cm 0.2cm 0.3cm,clip=true,width=0.45\textwidth]{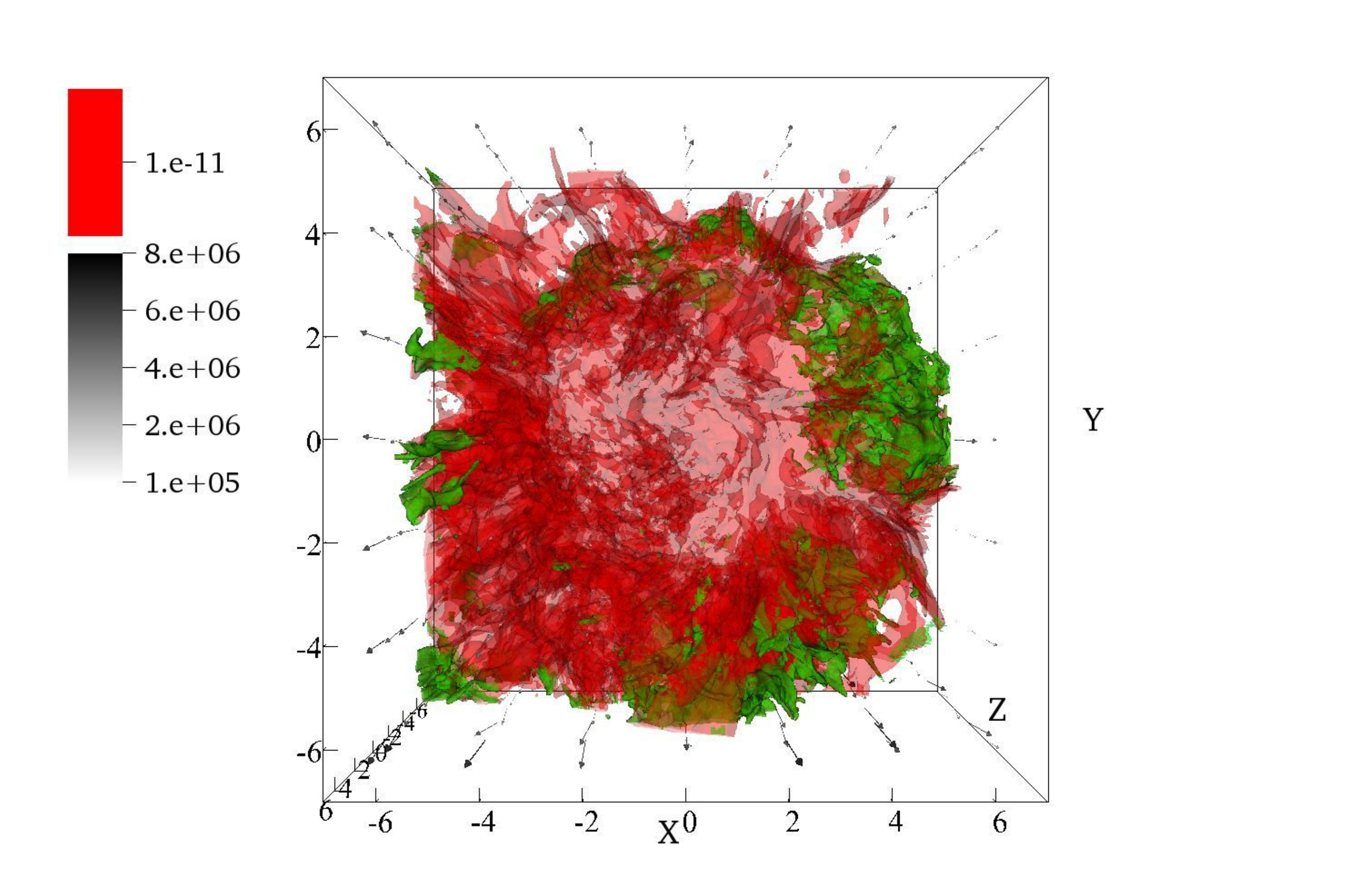}

\caption{ 3D maps of constant-density surfaces, each row for a different density surface, at the end of our fiducial run at $t=595$~day. 
The densities from top to bottom are $5\times 10^{-10},\;5\times 10^{-11}$ and $10^{-11} \g \cm^{-3}$, respectively.
Axes run from $-7\times 10^{13} \cm$ to $7\times 10^{13} \cm$.
In the left column the equatorial plane is along the line of sight in a horizontal plane through $z=0$, while in the right column the line of sight is perpendicular to the equatorial plane. The red color is the entire constant density surface, while the green color depicts unbound material, i.e., material which its sum of the kinetic, gravitational and internal energy is positive. The arrows depict velocity vectors, with the magnitude of the velocity proportional to the length of the arrow and to the brightness of the arrow as given in the gray bar on the left of each panel, and in the range of $1-80 \km \s^{-1}$.
}
\label{fig:dens_3d}
\end{figure*}

To further present the morphology of the outflowing gas, in Fig. \ref{fig:dens_3d} we present constant-density surfaces at the end of the fiducial run. Each row is for a different density value, and the two columns are for two different viewing angles as explain in the caption.
The red color depicts the entire constant-density surface. The green areas are the unbound regions, i.e., have positive total energy. 
The green color areas further present the concentrated of ejected mass near the equatorial plane.  
% FFFFFFFFFFFFFFFFFFFFFFFFFFFFFFFFFFFFFFFFFFFFFFFFFFFFFF
% FFFFFFFFFFFFFFFFFFFFFFFFFFFFFFFFFFFFFFFFFFFFFFFFFFFFFFF

{{{{ We emphasize that even though the jet are launched in directions perpendicular to the equatorial plane, most of the mass is ejected from the system in the equatorial plane directions. We can see it in two ways. One, as mentioned before, the jets are diverted towards the equatorial plane by the dense envelope. The denser the area around the companion the stronger bending of the jets will be. The other way is to realize that the jets create high-temperature low-density bubbles which escapes buoyantly outward, i.e., in and near the equatorial plane. 
These bubbles push and entrain envelope gas outward. }}}}

{{{{ The angular momentum in our simulations is affected solely by the jets as they are launched with the momentary velocity of the secondary star, and as we do not include neither the secondary gravity that can spin up the primary, nor initial giant rotation. The bending of the jets and the bubbles motion in our simulations distribute the angular momentum among the envelope parts they interact with. }}}}

% ==========================
\subsection{Varying the parameters}
\label{subsec:others}
% ==========================

The parameter space of the type of simulations we perform is huge. Below we describe the results of a limited number of simulations with varying values of some parameters, {{{{{ all with the low resolution numerical grid. }}}}}

\begin{figure*}
\centering
\includegraphics[trim= 1cm 0.4cm 0.2cm 0.3cm,clip=true,width=0.9\textwidth]{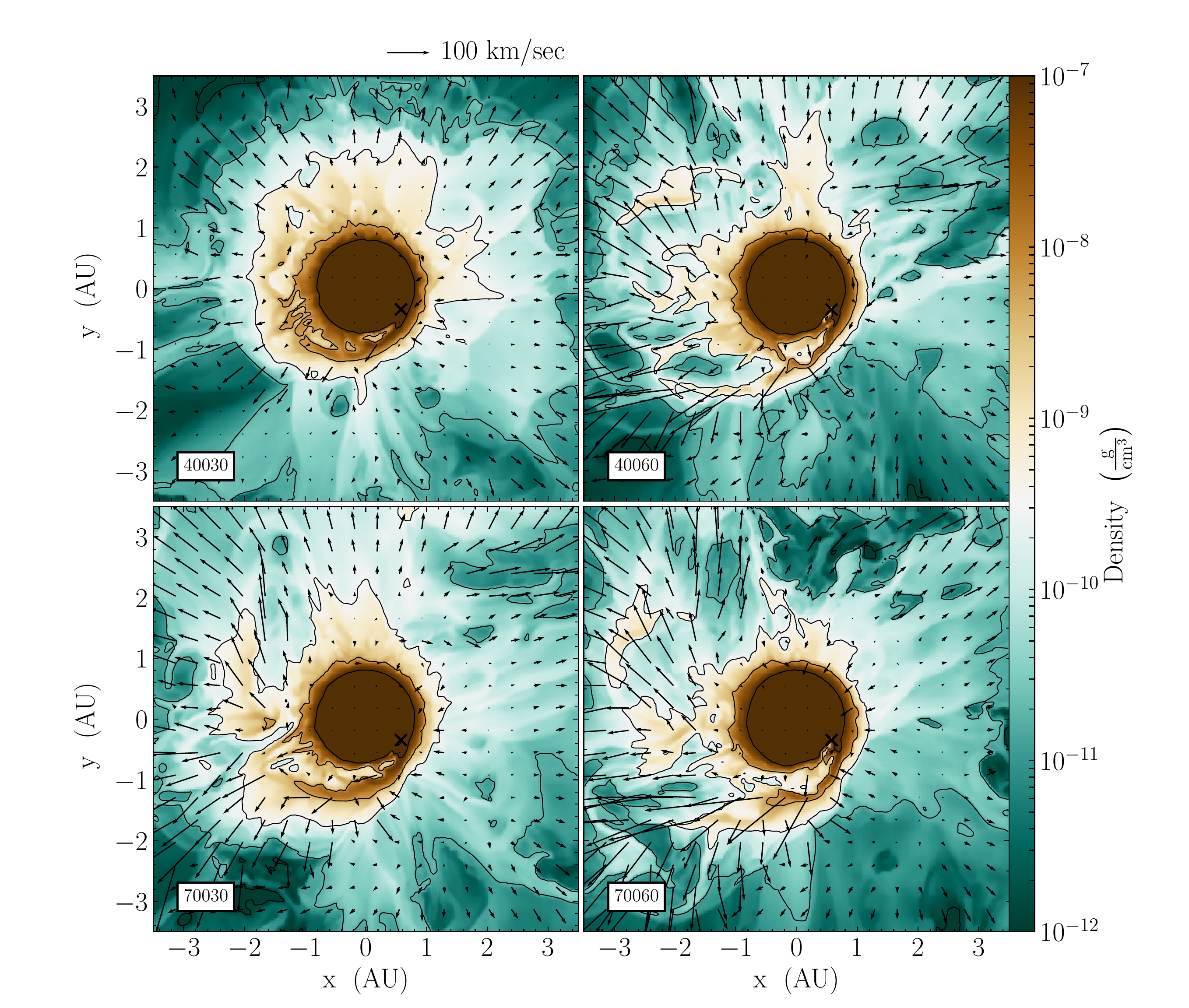} %l b r t
\caption{Density and velocity maps in the orbital plane $z = 0$ comparing four different runs after 589 days when the orbital separation has been reduced to $0.67\AU$. We label each panel by the value taken for the jets initial velocity relative to the secondary star $v_{\rm jet}$, and by the jets initial half-opening angle $\theta{\rm jet}$. The first three digits stand for the jet velocity in$\km \s^{-1}$, and the last two digits are the half opening angle in degrees. The upper left panel is the fiducial run whose flow structure is described in previous figures. The symbol 'X' in the outer part of the AGB star marks the location of the secondary star that rotates counterclockwise. The tail behind the secondary star in all panels show that as the secondary star dives into the envelope the jets remove mass from the outer part of the envelope. 
}
\label{fig:panel_varying_dens}
\end{figure*}

In Fig. \ref{fig:panel_varying_dens} we present density and velocity maps in the equatorial plane at the end of the simulations, when the orbital separation has been reduced to $0.67\AU$, for four simulations differ in the jet velocity and jet half-opening angle. The left upper panel shows the equatorial of a run where we set the jet velocity to be $v_{\rm jet}= 400 \km \s^{-1}$ and the half-opening angle to be $\theta_{\rm jet} = 30^\circ$. As we noted in subsections \ref{subsec:spiral} and \ref{subsec:outflow}, the structure of the flow has a spiral shape and the envelope is inflated outside the secondary orbit. The upper-right panel corresponds to a simulation with the same jet velocity and outflow mass rate but for wider jets of $\theta_{\rm jet}=60^\circ$. The spiral shape is more pronounced in this case, the ejected envelope flows faster, and more pronounced clumps of gas are ripped from the envelope (seen on the left part of the panel). The lower-left panel show the flow for the case of $v_{\rm jet}=700 \km \s^{-1}$ and $\theta_{\rm jet}=30^\circ$. As expected for a higher jet velocity which implies three times the jets' power, the envelope outer to the secondary orbit is more extended. In the lower-right panel the values are $v_{\rm jet}=700 \km \s^{-1}$ and $\theta_{\rm jet}=60^\circ$. As with the cases for a lower jet velocity, clumps of ejected envelope gas reach larger distances and velocities for the wider jets. In all panels we note the highly asymmetrical mass ejection. 
%FFFFFFFFFFFFFFFFFFFFFFFFFFFFFFFFFFFFFFFFFFFFFFFFFFFFFF
% FFFFFFFFFFFFFFFFFFFFFFFFFFFFFFFFFFFFFFFFFFFFFFFFFFFFFF

\begin{figure*}
\centering
\includegraphics[trim= 1cm 0.2cm 0.2cm 0.3cm,clip=true,width=0.9\textwidth]{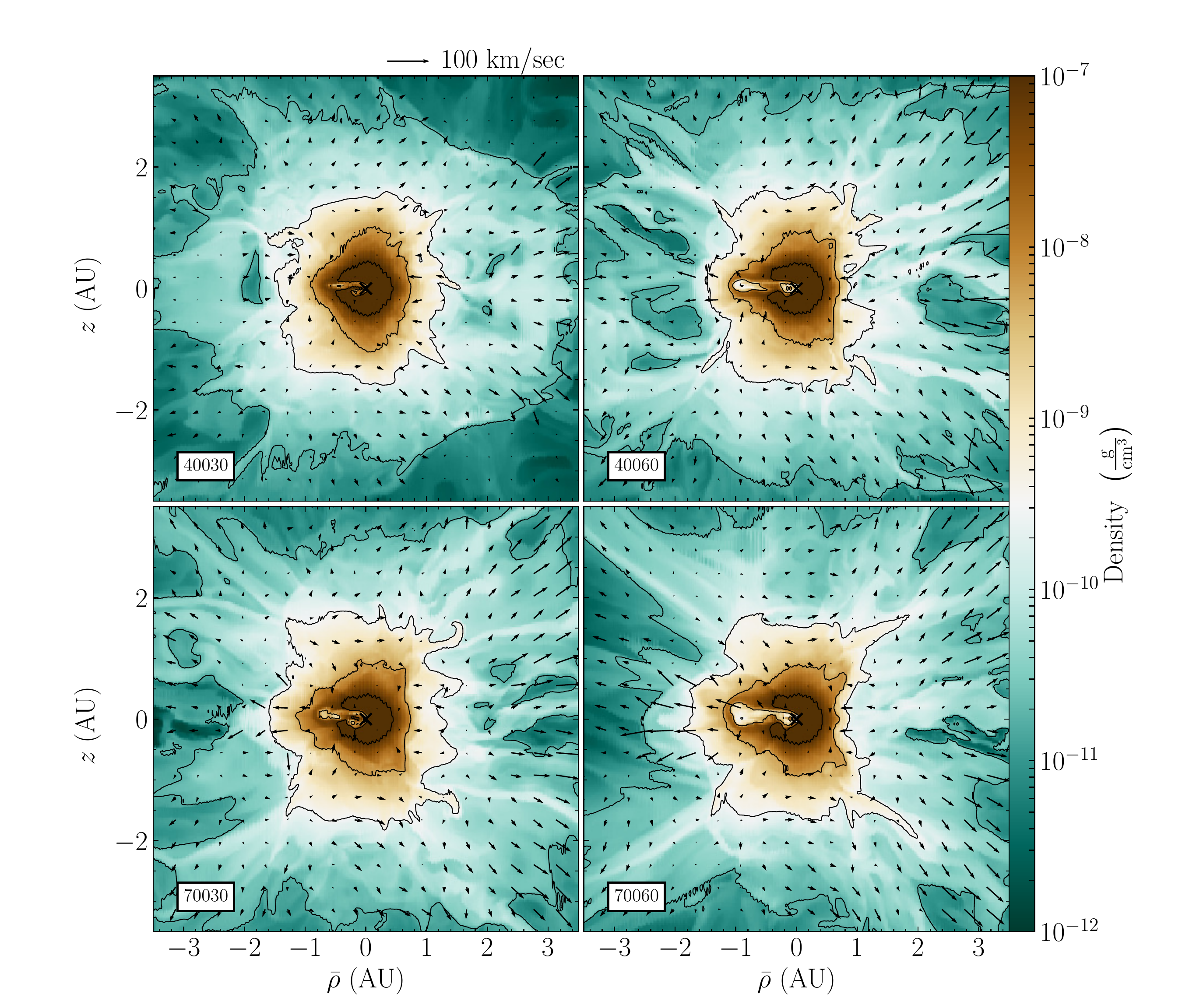} %l b r t
\caption{Like Fig. \ref{fig:panel_varying_dens}, but for the $\bar{\rho}-z$ plane that is perpendicular to the momentary radius vector of the secondary star and perpendicular to the equatorial plane. The center of the AGB star is at a distance of $0.67 \AU$ from this plane and projected onto the center of each panel.  These plots emphasize the highly asymmetrical mass ejection. 
}
\label{fig:panel_varying_tang}
\end{figure*}

In Fig. \ref{fig:panel_varying_tang} we compare the same four runs at the same time but present the density and velocity maps on a plane that is perpendicular to the momentary radius vector of the secondary star (the same plane that is shown in the lower panels in Fig. \ref{fig:panel_40030_end_2}). Asymmetrical pattern emanates from all the runs. The narrow trail behind the jets widens and escapes to larger distances with higher jet velocity and larger jet opening angle. The low density elongated structure starting from the center and extending to the left in each panel is the jets' material. Higher energy jets more easily escape from the envelope, and hence the higher energy simulations enter a common envelope phase at a later time.
%FFFFFFFFFFFFFFFFFFFFFFFFFFFFFFFFFFFFFFFFFFFFFFFFFFFFFF
% FFFFFFFFFFFFFFFFFFFFFFFFFFFFFFFFFFFFFFFFFFFFFFFFFFFFFF

\begin{figure*}
\centering
\includegraphics[trim= 1cm 0.2cm 0.2cm 0.3cm,clip=true,width=0.9\textwidth]{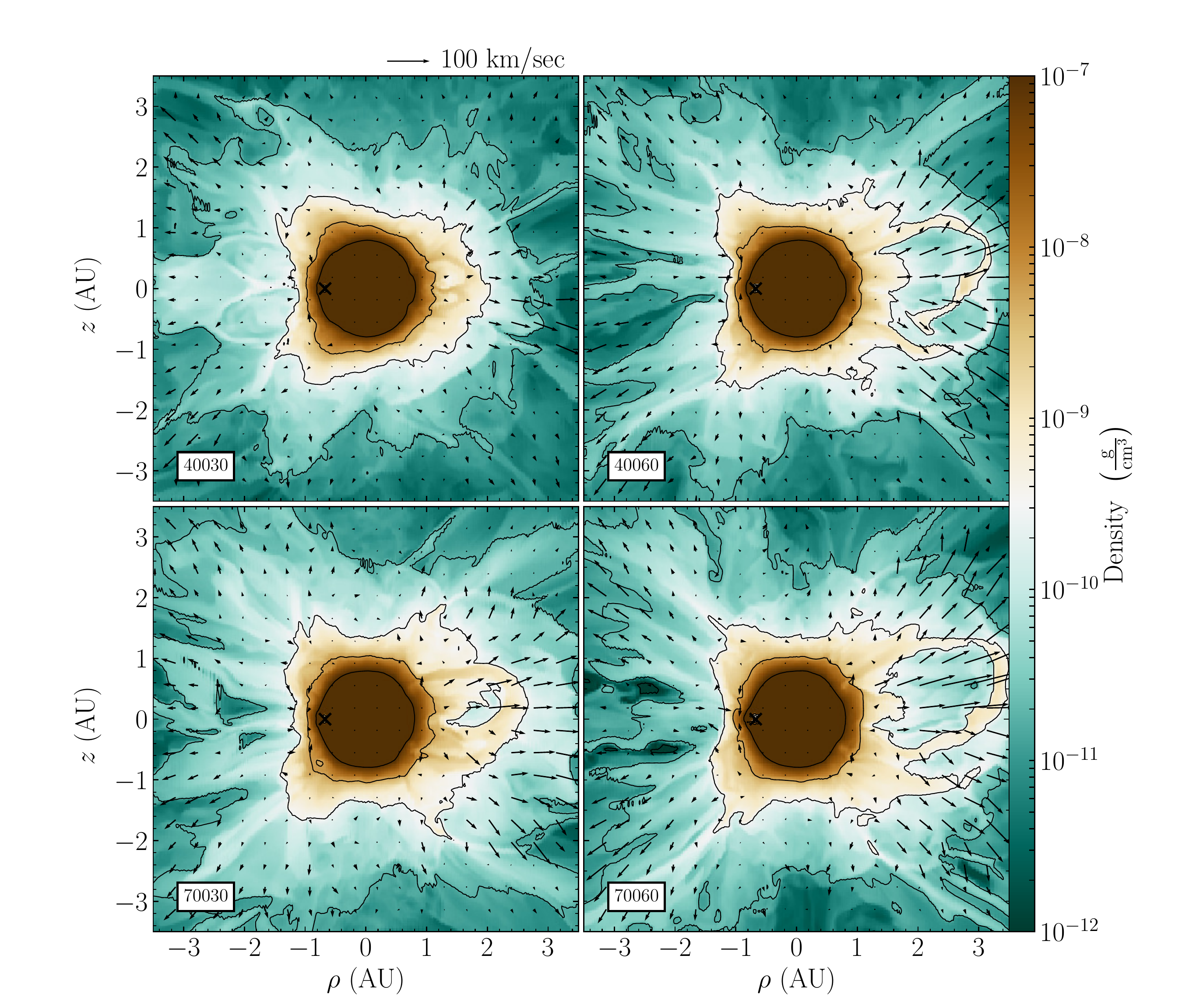} %l b r t
\caption{Like Fig. \ref{fig:panel_varying_dens}, but for the $\rho-z$ meridional plane that contains the center of the giant and the momentary location of the secondary star. Again, the asymmetrical mass ejection is clearly seen. 
}
\label{fig:panel_varying_center}
\end{figure*}

Fig. \ref{fig:panel_varying_center} shows the meridional plane that contains the momentary position of the secondary star and the center of the AGB star for the same runs and at the same time as in the previous two figures. Loops are seen on the right side of each panel. They are larger for higher velocity and higher jets' opening angles. These loops are part of the spiral outflow.  %FFFFFFFFFFFFFFFFFFFFFFFFFFFFFFFFFFFFFFFFFFFFFFFFFFFFFF
% FFFFFFFFFFFFFFFFFFFFFFFFFFFFFFFFFFFFFFFFFFFFFFFFFFFFFF

\begin{figure*}
\centering
\includegraphics[trim= 1cm 0.2cm 0.2cm 0.3cm,clip=true,width=0.9\textwidth]{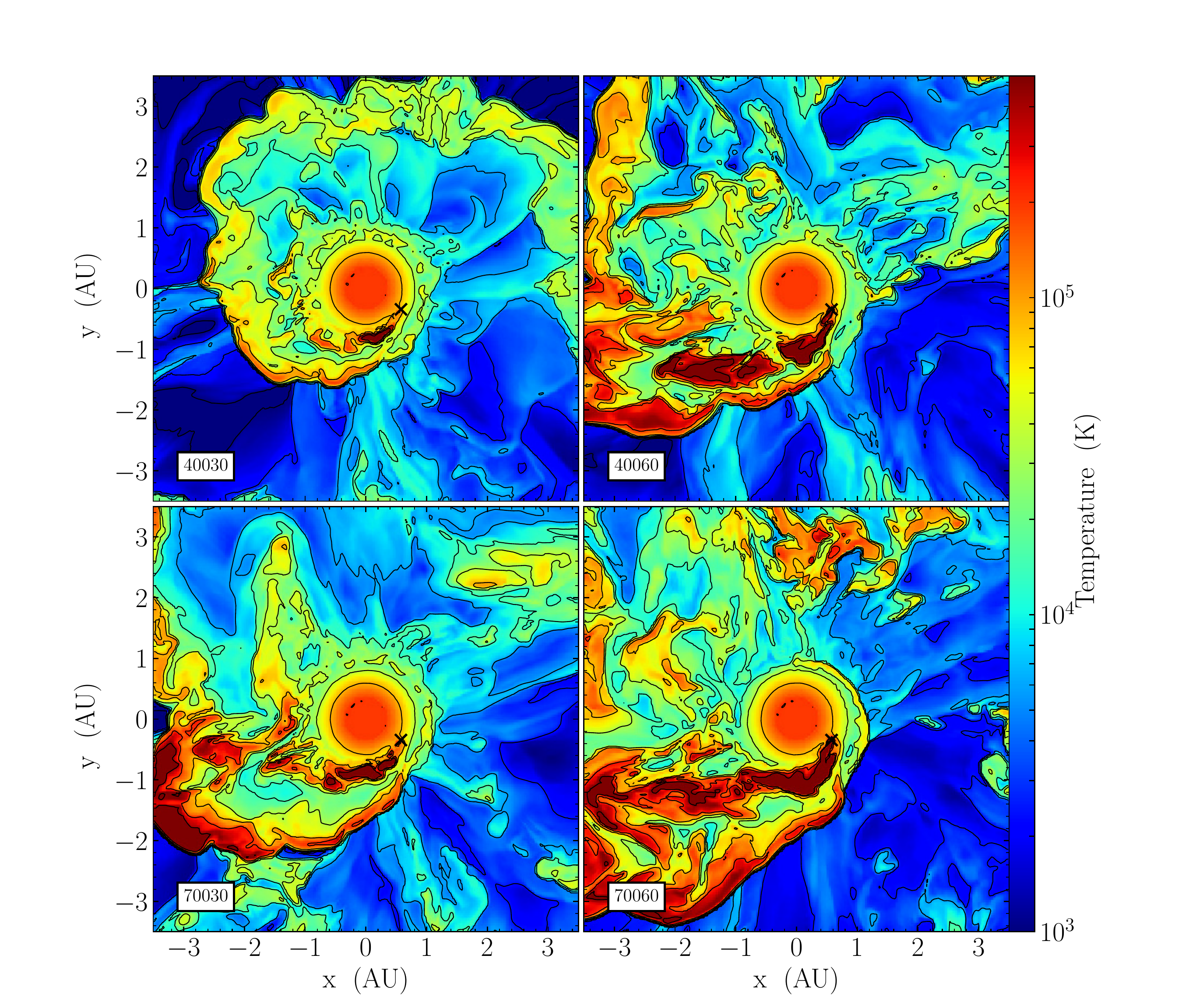} %l b r t
\caption{Like Fig. \ref{fig:panel_varying_dens} but showing the temperature maps. Wider jets result in wider tails trailing behind the secondary star. Jets with higher velocity expel the envelope mass more vigorously.
}
\label{fig:panel_varying_temp}
\end{figure*}

We present temperature maps from the same runs and at the same time as in previous three figures in Fig. \ref{fig:panel_varying_temp}. The post-shock material of the ejected gas forms a spiral pattern trailing the secondary star. A higher velocity yields larger regions of high temperature gas, and a wider opening angle of the jets results in a more complicated pattern and a wider spread of the spiral pattern. As discussed in relation to Fig. \ref{fig:temperature12}, the hot gas suffers a radiatively cooling in addition to the adiabatic cooling, and might lead to an ILOT event. 
%FFFFFFFFFFFFFFFFFFFFFFFFFFFFFFFFFFFFFFFFFFFFFFFFFFFFFF
% FFFFFFFFFFFFFFFFFFFFFFFFFFFFFFFFFFFFFFFFFFFFFFFFFFFFFF
  
\begin{figure*}
\centering
\includegraphics[trim= 4cm 0.2cm 0.2cm 0.3cm,clip=true,height=0.85\textheight]{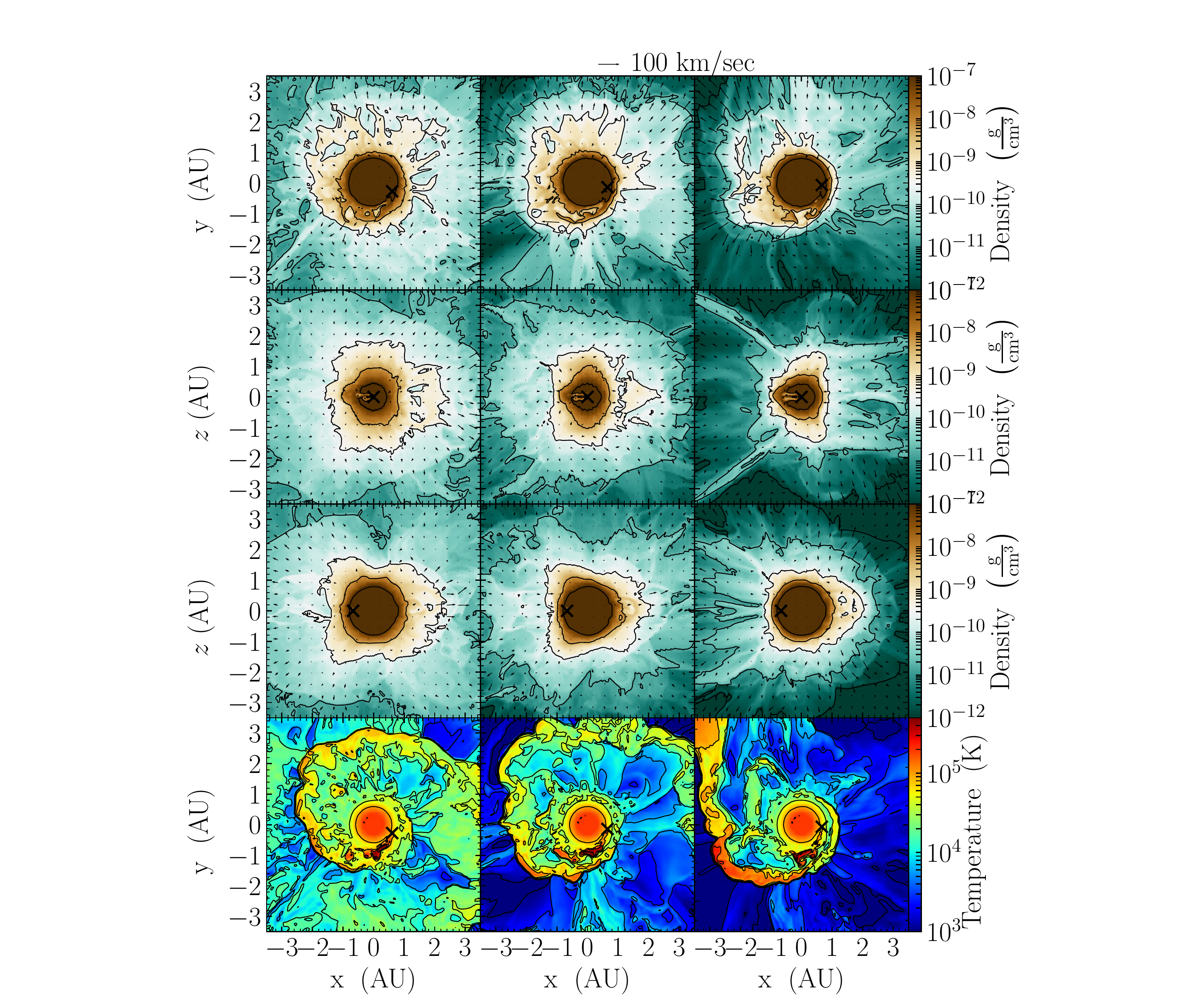} %l b r t
\caption{Comparison of the hydrodynamic properties of three runs at a final orbital separation of $a=0.67 \AU$, that are differ in the in-spiral time $t_{\rm sp}$. All the other parameters are as in the fiducial run. The left column is for $t_{\rm sp}=1190~$day, the middle column contains plots of the fiducial run for which $t_{\rm sp}=595~$day, and the right column presents plots for a simulation with 
$t_{\rm sp}=298~$day. 
The top three rows show density and velocity maps in the orbital plane z = 0 (top row), in a plane that is perpendicular to the momentary radius vector of the secondary star and perpendicular to the equatorial plane (second row), and in the meridional plane that contains the center of the giant and the momentary location of the secondary star (third row). The bottom row shows temperature maps in the orbital plane.
}
\label{fig:panel_varying_vr}
\end{figure*}

We also study the effect of the spiraling-in time. In Fig. \ref{fig:panel_varying_vr} we show plots of three runs differ in the spiral-in time $t_{\rm sp}$ from $a_0=1 \AU$ to $a=0.67 \AU$. The left column is for $t_{\rm sp}=1190\days$ equals to 6 Keplerian orbits around the AGB surface, but due to the spiraling-in motion the secondary completes almost 8 rounds. The middle column is for $t_{\rm sp}=595 \days$, our fiducial run. In the right column $t_{\rm sp}=298\days$ equals to 1.5 Keplerian orbits around the AGB surface, but with the spiraling-in motion the companion completes around 2 rounds. In all the runs we set $v_{\rm jet}=400 \km \s^{-1}$ and $\theta_{\rm jet}=30^\circ$, and we end the simulation when the orbital separation has shrank to $a=0.67 \AU$. 
In the figure we present density and velocity maps in three planes, the equatorial plane (top), the perpendicular plane (second row), and the meridional plane (third row), as explained in the caption. In the bottom row we present the temperature maps in the equatorial plane. 
%FFFFFFFFFFFFFFFFFFFFFFFFFFFFFFFFFFFFFFFFFFFFFFFFFFFFFF
% FFFFFFFFFFFFFFFFFFFFFFFFFFFFFFFFFFFFFFFFFFFFFFFFFFFFFF

As expected, when the spiraling-in time scale increases and more energy is injected in the jets by the time the secondary star reaches $a=0.67 \AU$, the envelope suffers more pronounced inflation. 
Also expected is that for a longer in-spiral time the geometry of the envelope ejection will be more isotropic. The temperature maps, for example, show the more sharply appearing spiral shape in the rapid plunge-in case relative to the slower spiraling-in where the arms are more extended. 

The main conclusion from the six different simulations we have described above in figs. \ref{fig:panel_varying_dens} - \ref{fig:panel_varying_vr}
is that in all cases the jets manage to inflate the envelope and eject some envelope gas. {{{{We imply from this conclusion that the GEE is a promising process to help in removing the envelope, as long as the secondary star launches energetic jets. }}}

% ==========================================================
\section{SUMMARY}
\label{sec:summary}
% ==========================================================

We simulated a secondary star that spirals-in from the surface of an AGB star at an orbital separation of $a_0=R_{\rm AGB}=1 \AU$, to a radius of $a=0.67 \AU$. We assumed that the secondary star launches jets continuously as it spirals-in, with a power that is several percents of that expected from the Bondy-Hoyle-Lyttleton accretion rate. The jet velocity is about equal to the escape velocity from the secondary star. Our goal is to explore the effect of jets, without yet introducing the orbital energy and the rotation of the envelope. 
 
 We revealed the structure of the outflow that results solely from the jets (figs. \ref{fig:density12}-\ref{fig:panel_40030_end}). The jets inflate the envelope outside the orbit, and eject some of the mass. The secondary star spiraled-in through an envelope layer with a mass of $0.37 M_\odot$, and ejected $0.018 M_\odot$. The very low ejected mass fraction is explained because the gravitational energy of the binary system and initial envelope rotation were not included. More efficient mass removal occurs when the secondary star orbits the outskirts of the envelope. 
 
Due to the bending of the jets by the AGB envelope, the mass outflow concentrates around the equatorial plane. The geometry is of an outflowing spiral pattern trailing the secondary star. Overall the flow is highly asymmetric and clumpy. Although most of the mass outflows from near the equatorial plane, the highest velocity of the outflow is at mid-latitude (fig. \ref{fig:mass_angle}). 

We examined the parameter space with additional five runs differ in the jets' velocity and the jets' half-opening angle (Figs. \ref{fig:outflow_map}-\ref{fig:panel_varying_temp}), and in the in-spiral time 
(Fig. \ref{fig:panel_varying_vr}). We find outflow in all cases. {{{{ We conclude that the GEE is a promising process. In some cases it might substantially postpone and even prevent the spiraling-in process. }}}}
However, to determine whether the jets play a significant role, the gravitational energy of the binary system and the spinning of the AGB envelope must be included. These are the tasks of future studies.

\section*{Acknowledgments}
{{{{ We thank an anonymous referee for comments that improved the presentation of our results. }}}}
This work was supported by the Cy-Tera Project, which is co-funded by the European Regional Development Fund and the Republic of Cyprus through the Research Promotion Foundation. 
{{{{{We thank Thekla Loizou from the Cyprus HPC Facility for her kind technical assistance. FLASH was developed
largely by the DOE-supported ASC/Alliances Center for Astrophysical
Thermonuclear Flashes at the University of Chicago. In producing the images in this paper we used VisIt which is supported by the Department of Energy with funding from the Advanced Simulation and Computing Program and the Scientific Discovery through Advanced Computing Program. This work also required the use and integration of a Python package for astronomy, yt (http://yt-project.org, \citealt{Turk2011}). }}}}}
We acknowledge support from the Israel Science Foundation and a grant from the Asher Space Research Institute at the Technion. 

%\footnotesize

\label{lastpage}
\end{document}